\documentclass[sigconf, nonacm]{acmart}
\graphicspath{{images/}}
\usepackage[a-1b]{pdfx} 
\usepackage{array}
\usepackage{amsmath}
\usepackage{algorithmic}
\usepackage[ruled,vlined]{algorithm2e}
\usepackage{pgfplots,pgfplotstable}  
\usepackage{booktabs}
\usepackage{balance}
\usepackage{multirow}
\usepackage{subcaption}
\pgfplotsset{compat=newest}
\usepackage{balance}

\usepackage{xcolor}

\newcommand{\tinyskip}{\vspace{1pt}}
\newcommand{\mypar}[1]{\tinyskip\tinyskip\noindent\textbf{#1.}\xspace}
\newcommand\vldbdoi{10.14778/3554821.3554840}
\newcommand\vldbpages{3509 - 3521}
\newcommand\vldbvolume{15}
\newcommand\vldbissue{12}
\newcommand\vldbyear{2022}
\newcommand\vldbauthors{\authors}
\newcommand\vldbtitle{\shorttitle} 
\newcommand\vldbavailabilityurl{}
\newcommand\vldbpagestyle{empty} 
\newcommand{\eat}[1]{}
\newcommand{\sysname}{Doppler\xspace}

\newcommand{\pp}{price-performance\xspace}

\newcounter{enum}

\begin{document}
%\title{\sysname: A Framework for Evaluating Price Performance Tradeoffs in Migrating SQL Workloads to the Cloud}
\title{\sysname: Automated SKU Recommendation in Migrating SQL Workloads to the Cloud}
%other titles? ideas: migration, on-prem, cloud, Azure SQL, price performance, elastic strategy 
%migration support system.. ?

%%
%% The "author" command and its associated commands are used to define the authors and their affiliations.
\author{Joyce Cahoon}
\affiliation{%
	\institution{Microsoft}
	\streetaddress{One Microsoft Way}
	\city{Redmond}
	\state{WA}
	\postcode{98052}
}
\email{jcahoon@microsoft.com}
%\authornote{Corresponding author.}
\authornote{Authors contributed equally.}

\author{Wenjing Wang}
\affiliation{%
	\institution{Microsoft}
	\streetaddress{One Microsoft Way}
	\city{Redmond}
	\state{WA}
	\postcode{98052}
}
\email{wenjwang@microsoft.com}
\authornotemark[1]

\author{Yiwen Zhu}
\affiliation{%
	\institution{Microsoft}
	\streetaddress{One Microsoft Way}
	\city{Redmond}
	\state{WA}
	\postcode{98052}
}
\email{yiwzh@microsoft.com}
\authornotemark[1]

\author{Katherine Lin}
\affiliation{%
	\institution{Microsoft}
	\streetaddress{One Microsoft Way}
	\city{Redmond}
	\state{WA}
	\postcode{98052}
}
\email{katlin@microsoft.com}

\author{Sean Liu}
\affiliation{%
	\institution{Microsoft}
	\streetaddress{One Microsoft Way}
	\city{Redmond}
	\state{WA}
	\postcode{98052}
}
\email{seliu@microsoft.com}

\author{Raymond Truong}
\affiliation{%
	\institution{Microsoft}
	\streetaddress{One Microsoft Way}
	\city{Redmond}
	\state{WA}
	\postcode{98052}
}
\email{ratruong@microsoft.com}

\author{Neetu Singh}
\affiliation{%
	\institution{Microsoft}
	\streetaddress{One Microsoft Way}
	\city{Redmond}
	\state{WA}
	\postcode{98052}
}
\email{nesin@microsoft.com}

\author{Chengcheng Wan}
\affiliation{%
	\institution{University of Chicago}
	\streetaddress{One Microsoft Way}
	\city{Chicago}
	\state{IL}
	\postcode{98052}
}
\email{cwan@uchicago.edu}
\authornote{Work done while interning at Microsoft.}

\author{Alexandra Ciortea}
\affiliation{%
	\institution{Microsoft}
	\streetaddress{One Microsoft Way}
	\city{Redmond}
	\state{WA}
	\postcode{98052}
}
\email{aciortea@microsoft.com}

\author{Sreraman Narasimhan}
\affiliation{%
	\institution{Microsoft}
	\streetaddress{One Microsoft Way}
	\city{Redmond}
	\state{WA}
	\postcode{98052}
}
\email{sreraman@microsoft.com}

%\author{Pratyush Rawat}
%\affiliation{%
%	\institution{Microsoft}
%	\streetaddress{One Microsoft Way}
%	\city{Redmond}
%	\state{WA}
%	\postcode{98052}
%}
%\email{pratraw@microsoft.com}
%
%
%\author{Haitao Song}
%\affiliation{%
%	\institution{DocuSign}
%	\streetaddress{One Microsoft Way}
%	\city{Redmond}
%	\state{WA}
%	\postcode{98052}
%}
%\email{haitao.song@gmail.com}

\author{Subru Krishnan}
\affiliation{%
	\institution{Microsoft}
	\streetaddress{One Microsoft Way}
	\city{Mountain View}
	\state{CA}
	\postcode{98052}
}
\email{subru@microsoft.com}

% \newcommand{\autspace}{~~~~~~~~~~}
% \author{
% 	Joyce Cahoon$^1$ \autspace Yiwen Zhu$^1$ \autspace Wenjing Wang$^1$ \autspace Katherine Lin$^1$ \autspace Subramaniam Venkatraman$^1$}
% \author{
% 	\autspace Sean Liu$^1$ \autspace Raymond Truong$^1$}
% \affiliation{ 
% 	\institution{$^1$Microsoft, firstname.lastname@microsoft.com}
%!TEX root = main.tex

\begin{abstract}

Selecting the optimal cloud target to migrate SQL estates from on-premises to the cloud remains a challenge. Current solutions are not only time-consuming and error-prone, requiring significant user input, but also fail to provide appropriate recommendations. We present \sysname, a scalable recommendation engine that provides right-sized Azure SQL Platform-as-a-Service (PaaS) recommendations without requiring access to sensitive customer data and queries. 
%In the journey of migrating SQL instances from on-premises to cloud, identifying the appropriate SKU is a major challenging that requires significant input from database administrators or field engineers, which is time-consuming and error-prone. Moreover, customer workload data is often inaccessible due to a higher level of privacy and security concerns.
%In this paper, we present \sysname, a scalable SKU recommendation engine that provides right-sized Azure SQL PaaS recommendations without accessing customer query/data. 
%\sysname provides a consistent framework for mapping on-premise workloads to appropriate PaaS cloud targets by introducing a novel \pp methodology that allows customers to get a personalized rank of relevant cloud targets. 
\sysname introduces a novel \pp methodology that allows customers to get a personalized rank of relevant cloud targets solely based on low-level resource statistics, such as latency and memory usage. 
%This rank is obtained solely from the input of low-level resource statistics, such as latency and memory usage. 
\sysname supplements this rank with internal knowledge of Azure customer behavior to help guide new migration customers towards one optimal target. Experimental results over a 9-month period from prospective and existing customers indicate that \sysname can identify optimal targets and adapt to changes in customer workloads. 
It has also found cost-saving opportunities among over-provisioned cloud customers, without compromising on capacity or other requirements. \sysname has been integrated and released in the Azure Data Migration Assistant v5.5, which receives hundreds of assessment requests daily.

%matches between SQL instance SKUs to the customers' workload and capacity requirements via a combination of traditional price-performance methodology and customer profiling techniques, 

%leveraging the massive (aggregated) data from existing customers. The price-performance approach provides a personalized rank of all the relevant on-cloud SKUs that a customer can migrate towards, while the customer profiler leverages internal knowledge of Azure customer behavior to support the decision making. 
%in order to provide better customized solutions for new customers. 
%Experimental results using telemetry from prospective customers, supplemented by richer telemetry from customers already in the cloud, indicate that \sysname can detect and adapt to changes in customer workloads. 
%simply by leveraging data collected from performance counters. 
%It has also identified cost-saving opportunities among existing cloud customers, 
%without compromising on capacity or other requirements. \sysname is now released as one of the SKU recommendation engines integrated in the Azure Data Migration Assistant v5.5~\cite{dma}, which receives hundreds of assessment requests daily. 

%We introduce a novel price-performance framework to contribute to the growing knowledge and set of tools available to migrate existing systems to the cloud. Since migration is a highly intricate process, we share the process of migrating legacy SQL customers to Azure SQL PaaS in order to help build best practices and methods in ... 
\end{abstract}

%The market opportunity for migrating legacy SQL customers to Azure SQL has been estimated at 8 billion. There exist many methods that support migrating legacy systems to the cloud, from generic surveys to complex machine learning algorithms; yet, despite this rich ecosystem of solutions, there is not yet one universal, automated solution that cloud providers rely on to migrate these customers to the cloud. For one, customers face a bewildering array of destination SKU choices during migration. This paper focuses on making accurate, personalized and interpretable SKU recommendations to help increase customers propensity to migrate and reduce post-migration friction. We introduce a novel E2E SKU recommendation framework based on traditional price-performance methodology to make recommendations for migrating on-prem customers to Azure SQL PaaS. This approach will be released with the next version of Azure Data Migration Assistant, a tool for SKU recommendation and migration assessment. 
\maketitle

\pagestyle{\vldbpagestyle}
\begingroup\small\noindent\raggedright\textbf{PVLDB Reference Format:}\\
\vldbauthors. \vldbtitle. PVLDB, \vldbvolume(\vldbissue): \vldbpages, \vldbyear.\\
\href{https://doi.org/\vldbdoi}{doi:\vldbdoi}
\endgroup
\begingroup
\renewcommand\thefootnote{}\footnote{\noindent
	This work is licensed under the Creative Commons BY-NC-ND 4.0 International License. Visit \url{https://creativecommons.org/licenses/by-nc-nd/4.0/} to view a copy of this license. For any use beyond those covered by this license, obtain permission by emailing \href{mailto:info@vldb.org}{info@vldb.org}. Copyright is held by the owner/author(s). Publication rights licensed to the VLDB Endowment. \\
	\raggedright Proceedings of the VLDB Endowment, Vol. \vldbvolume, No. \vldbissue\ %
	ISSN 2150-8097. \\
	\href{https://doi.org/\vldbdoi}{doi:\vldbdoi} \\
}\addtocounter{footnote}{-1}\endgroup

%\ifdefempty{\vldbavailabilityurl}{}{
%\vspace{.3cm}
%\begingroup\small\noindent\raggedright\textbf{PVLDB Artifact Availability:}\\
%The source code, data, and/or other artifacts have been made available at \url{\vldbavailabilityurl}.
%\endgroup

\section{Introduction}
\label{sec:intro}

% PROBLEM STATEMENT: MIGRATION IS HARD 
The complexities of migrating SQL estates from an on-premise data platform to the cloud cannot be understated. An extensive assessment process first takes place to evaluate what cloud targets can accommodate existing workloads. Identifying the optimal targets remains a challenge, as it not only involves understanding the compute resources required to handle customer workloads, but also involves analyzing the legacy systems as a whole---its source code, binaries, configuration files, and execution traces---to ensure feature parity and check for compatibility issues that may arise in the migration process. If (or when) optimal targets are found, multiple stakeholders must then be mobilized to execute the migration to ensure data and their applications can be ported and remains intact and secure during the process. Without proper planning, migration can lead to degraded workload performance and higher costs. 

\begin{figure}[t]
	\centering
	\includegraphics[width=\columnwidth]{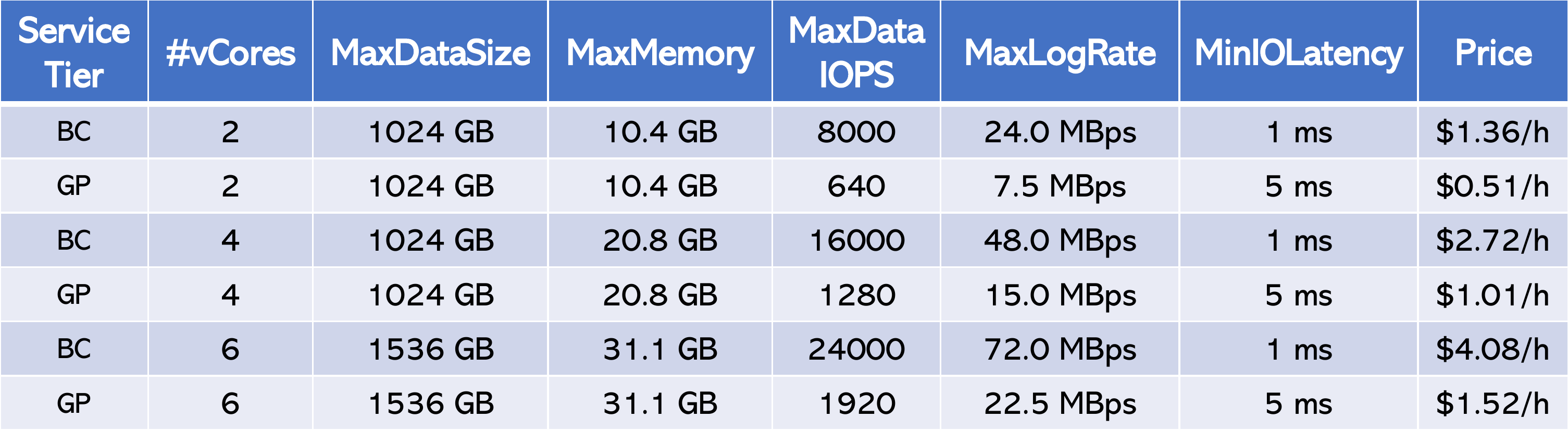}
%	\vspace{-0.6cm}
	\caption{Examples of 6 Azure SQL SKU offerings~\citep{DBResourceLimits, MIResourceLimits,price}.}\label{fig:azure_db_choices}
\vspace{-0.6cm}
\end{figure}

%Since the total addressable market (TAM) for migrating on-premise data platforms to PaaS offerings in the cloud is estimated at \$8.1 billion,
Since the DBMS market is estimated at \$64.8 billion~\cite{gartner1}, and it is predicted that 75\% of all databases will be migrated (or deployed) from a cloud platform~\cite{gartner2}, the total addressable market (TAM) for migrating on-premise data platforms to PaaS offerings is large. Providers have funneled resources to provide an ecosystem of solutions to ease the migration process. 
Current solutions range from increasing the diversity of cloud targets, also known as Stock Keeping Units (SKU), to automating various steps of the migration process to meet customer preferences in terms of budget and performance. Figure~\ref{fig:azure_db_choices} illustrates a few examples of different Azure SQL Database SKU offerings, but this only accounts for about 2\% of all the possible SKUs. They are architected to cater to a variety of customer workload requirements in terms of transaction rates, latency, throughput, CPU, memory, and storage.

Many tools exist to assist the process of selecting the right cloud target~\cite{kopaneli2015model,kang2010cloudle}. Despite the utility of these strategies, they require significant input from the customer, and the final recommendations may still be inappropriate. 
% Several surveys \citep{gholami2016cloud, razavian2015systematic, jamshidi2013cloud, khadka2013legacy, lane2011process, wei2010service} highlight the complexity of this decision-making point and cite tools like the Cloud Target Selection (CTS)~\cite{kopaneli2015model} and Cloudle~\cite{kang2010cloudle} that support migrating legacy applications to the cloud. CTS is a large question catalogue that aids new migration customers in choosing the right provider and cloud SKU; Cloudle is similar, but also allows customers to specify additional functional, technical and cost requirements. 
As a result, cloud providers default to manual SKU selection as the de facto standard as the decision support systems proposed (e.g.,~\cite{kang2010cloudle, beserra2012cloudstep, kopaneli2015model}) are too hard to use and are difficult to scale.
%Our field experiences have shown that proper SKU selection remains a problem as of 125k databases assessed per year, only 60k proceed with the SKU selections and migrate to Azure SQL.
% Our field experiences have shown that proper SKU selection remains a problem as hundreds of thousands of databases are assessed per year, and a fraction proceeds to migrate.
Our field experiences have shown that proper SKU selection remains a problem that can take up to 60\% of the total migration journey. Hundreds of thousands of databases are assessed per year, but only a fraction proceeds to migrate.
%alone accounts for more than \ww{TBD}\%, or an average of \ww{ can take up to 4 weeks}, in the total migration journey \ww{which can take months to finish}.  
There does not yet exist a straightforward way for customers to exit their existing data centers and land on an optimal cloud SKU~\cite{gholami2016cloud,girish2014survey,alkhalil2017decision}. 
As a result, a large segment of customers with SQL estates on-premise abstain from moving their workloads to the cloud because of the risks that come with the migration process~\citep{andrikopoulos2014design, andrikopoulos2013supporting, klems2008clouds,gholami2016cloud,girish2014survey,alkhalil2017decision}.
% which includes the possible lack of business continuity, the degradation in performance, and the increase cost in operation. 
% These issues arise because there is no set recipe for migrating customers to the cloud.

In this paper, we develop a data-driven SKU recommendation tool that automates the PaaS cloud target selection process and addresses the following challenges:

\mypar{Privacy concerns}
The most accurate approach for cloud SKU selection is workload replay.
% ~\cite{stitcher}. 
It requires accessing user data and query history and replaying the exact same operations on a new cloud SKU as the customer had executed in an on-premise data platform. However, this approach is not practical as there are few customers that are comfortable with providing such unrestricted access. Recent work has demonstrated attempts to replay anonymized data, but this methodology still requires access to the underlying raw user files to do so~\cite{deep2021diametrics}.

\mypar{Too many SKU options}
Microsoft Azure alone has over 200 different PaaS cloud SKUs. 
% They are architected to cater to a variety of customer workload requirements in terms of transaction rates, latency, throughput, CPU, memory, and storage. 
These SKUs are primarily segmented by deployment type, Azure SQL Database (DB) or Managed Instance (MI), and service tiers, i.e., General Purpose (GP) or Business Critical (BC)---the latter of which offers higher resilience to failure. The sheer volume of options often leads to decision paralysis, increasing the inertia among customers to modernize and migrate to the cloud.

\mypar{Lack of transparency in SKU choice}
While enough data now exists to leverage black-box machine learning (ML) methods for SKU selection, these approaches lack interpretability. This includes a previous internal attempt to do automated SKU recommendation by training neural nets on workload characteristics, but the cloud targets selected from these models cannot be explained. As the risk associated with migration is high, customers need to understand why a specific SKU choice is made.
 
% Microsoft Azure alone has over 200 different PaaS cloud SKUs. They are architected to cater to a variety of customer workload requirements in terms of transaction rates, latency, throughput, CPU, memory, and storage. 
% These SKUs are primarily segmented by deployment type, Azure SQL Database (DB) or Managed Instance (MI), and service tiers, General Purpose (GP) or Business Critical (BC)---the latter of which offers higher resilience to failure. Figure~\ref{fig:azure_db_choices} illustrates a few examples of different Azure SQL Database SKU offerings, but this only accounts for about 2\% of all the possible SKUs. 

%strict technical requirements for different applications often fail a one-size-fits-all model; 

\mypar{Lack of customized solutions}
Field engineers depend on intuition and migration experience to gauge a customers' workload requirements. For example, some applications can function under memory constraints, while others have no tolerance for hitting specific resource boundaries. Since migration assessment still depends on expert opinion and ad-hoc analyses, and customers vary greatly in their tolerance for these risks, e.g., some are willing to sacrifice on certain aspects of performance for increased cost savings \cite{andrikopoulos2013supporting}, it is difficult to develop a fully automated system that quantifies these differences in preference and captures this complex decision-making process by experienced engineers.

\subsubsection*{\textbf{Introduction to \sysname}}
%\subsection{Overview of \sysname}
% NOVELTY OF OUR APPROACH , need to make it sound less ADHOC
\sysname relies solely on customers' workload performance history, or low-level resource statistics, such as latency and memory usage, as input to make a data-driven SKU selection. We address the aforementioned challenges as follows:
%To address the above issues,we developed \textit{\sysname} that provides more accurate, explainable SKU recommendation results, and automates the assessment phase of the migration journey to Azure SQL.

\textit{\sysname relies only on resource consumption patterns to evaluate the suitability of SKUs.}
We circumvent the problem of accessing customer data and query history by characterizing workload behavior solely through resource consumption patterns. Exploratory analyses of various workloads (and their subsequent performance history) suggest that such low-level resource statistics are sufficient to capture differences in workload, and thus differences in resource demands.
% ~\cite{stitcher}. 
Our novel methodology addresses current challenges in privacy and security and avoids time-consuming processes, like replaying customer workloads~\cite{kopaneli2015model} and building complicated workload simulation platforms~\cite{saez2015performance}. It can be easily extended to accommodate additional performance dimensions outside of the current feature set.
%As opposed to simulating the workload or actuallyon all SKUs, \sysname focuses on the summarization of the resource consumption patterns using statistical methods to evaluate the performance. This novel characterization of workloads avoids the needs of accessing customer workload data, which involves a higher level of privacy and security concerns, or the needs of building complicated simulation platforms.

%\textit{\sysname provides an automated and easy-to-understand process for SKU recommendation.}
\textit{\sysname provides an automated and interpretable process for SKU recommendation.}
%We focus on developing an automated, model-driven approach to facilitate the assessment phase of customers' migration journey, and to provide more accurate, explainable SKU recommendation results to help migrate customers to Azure SQL. 
Existing migration tools concentrate on optimizing cost. \sysname presents a suite of optimal SKUs in the context of, not only cost, but also performance. 
\sysname summarizes the trade-offs between cost and performance for each SKU in the form of a \pp curve, an example of which is shown in Figure~\ref{fig:ex_weight2}. 
Our use of our customer workload's performance history allows us to judge how well each workload will perform on a new cloud SKU based on concrete metrics like compute, memory, latency and throughput. 
Since customers' appetite for risk varies, \sysname outputs a \pp curve to
%Since customers' appetite for risk changes, \sysname outputs a \pp curve (an example shown in Figure~\ref{fig:ex_weight}) to
rank how various SKUs fulfill performance needs, along with its subsequent impact on cost, to provide the customer 
%not only with clear reasoning behind optimal SKU selections, but also 
with the agency needed to make an informed decision.

%\yz{updated below a bit~}
%\textit{\sysname ``learns'' from existing cloud customers for whom the decisions are made by migration experts}. 
\textit{\sysname ``learns'' from previous optimal\footnote{Since there is no benchmark (method) that definitively marks a SKU as ``optimal,'' we developed our own from performance histories associated with successfully migrated customers that have fixed their SKU for at least 40 days. In this paper, our claims of ``optimal'' SKUs are established relative to successfully migrated Azure customers, excluding customers that are over-provisioned.} SKU choices vetted by migration experts}.
%and captures workload heterogeneity}. 
%Doppler learns from existing cloud customers to understand user preferences and provide personalized recommendations. 
Based on Azure cloud customers (who have fixed their chosen SKUs for at least 40 days), we segment the customers based on their performance footprint as a means of categorizing their workload behavior. 
This involves studying how their \pp curves change and observing what SKU they fix their workloads to based on the resulting \pp curves. % and resource utilization patterns. 
%Interestingly, customers do not always choose the cheapest SKU that fulfills their workload resource needs at 100\%. 
There are a large number of customers willing to negotiate on certain performance dimensions, like latency and throughput, to realize cost savings. 
%By generating \pp curves and measure what perf dimensions might be negotiable, we quantify preferences for each customer segment in terms of cost over performance. 
By generating price-performance curves and measuring what performance dimensions may be negotiable, we can quantify customer preferences in a way that aligns with how experts vet optimal cloud targets.
In other words, we introduce a data-driven strategy that automatically characterizes customers’ workload needs as shown in Figure~\ref{fig:doppler} in order to provide personalized SKU recommendations. % solutions based on each cudifferent workload profiles. %We translate this knowledge to help new migration customers, 
% By providing a reference point to our existing customer base, we guide new migration customers towards selecting a better (and cheaper) SKU.

%\yz{Through the customer profiling analysis, which is based on the resource usage pattern in each dimensions, we aim at embedding the knowledge from the experts on different levels of tolerance in various resource dimensions in the categorization process of workload patterns and mimic the previous engineering effort that leads to customized solution that captures different levels of tolerances for heterogeneous workloads. With \sysname, the selection process by experts can be democratized and any non-experts can leverage the tool for the recommendation easily.}

\textit{\sysname is an integral part of the Data Migration Assistant (DMA) tool~\cite{dma} and has been used by thousands of customers since its release.}
%to facilitate the end-to-end migration, which
%\change{It entails extensive due diligence on behalf of our field engineers and system integration partners: prepping inventories of our customers' on-prem infrastructure, evaluating the size of the SQL Server estate, matching it to available internal inventory, assessing the complexity of workloads and their downstream dependencies, and generating business reports to make the case for migration.}\cmt{not sure since we observe many customers can actually do their own assessement and migrations without intervention by using migration tool or some third party tool.}
We executed an experimental evaluation of \sysname over 9 months of customer data to verify that we can identify optimal cloud SKUs
% \footnote{Relative to successfully migrated Azure SQL PaaS customers that have fixed their SKU choice for at least 40 days.} 
with our framework.
Given that \sysname is able to generate the exact same SKU choice as 89.4\% and 96.7\% of successfully migrated SQL DB and MI customers, whose SKUs were vetted by migration experts, \sysname was subsequently implemented in Azure's DMA tool.
Every new Azure migration customer uses DMA to begin an assessment of their on-premise data servers. 
% Our DMA tool executes admin scripts to collect SQL instance performance counters, upon which \sysname consumes to make SKU recommendations. DMA also automatically detects compatibility issues that can impact database functionality on the recommended SKUs from \sysname. 
Given the ease of use and utility of this migration assistant, it has been utilized by hundreds of customers since its release in October 2021 (see Table~\ref{tab:dmausage}). 
%Since it takes, \ww{months} on average for field engineers to make one SKU recommendation, with \sysname, we have saved our partners at least \ww{XXX} hours in the migration assessment process. \jc{reduce months to hours whole migration process}

\begin{table}[t]
	\caption{DMA tool adoption since its release.}
	\label{tab:dmausage}
	\vspace{-2mm}
% 	\resizebox{\columnwidth}{!}{%
		\begin{tabular}{c p{0.22\linewidth} p{0.25\linewidth} p{0.25\linewidth}}
			\toprule
			Month  & Unique instances assessed  & Unique databases assessed    & Total recommendations generated  \\
			\midrule
			Oct-21  & 185 & 3,905 & 6,503  \\
			Nov-21  & 215 & 3,389 & 4,802        \\
			Dec-21 & 57 & 4,185 & 5,364    \\
			Jan-22 & 231 & 9,090 & 10,674    \\
			\bottomrule
		\end{tabular}
% 	}
% 	\vspace{-4mm}
\vspace{-.6cm}
\end{table}

\textit{\sysname provides a consistent framework for making SKU recommendations not only for new migration customers but also existing cloud customers.}
In leveraging Azure customers' historical performance profiles, we found that we were not just able to increase the accuracy of our SKU recommendation for migrating new customers' SQL estates from an on-premise data platform to the cloud, but also right-size SKUs for our existing customer base. In one recent customer survey, over a 30-day observation window, we found that 30\% of SQL databases consume 43\% or less of provisioned CPU resources, and that only 5\% of SQL databases reach the maximum provisioned CPU usage for more than 10\% of this study's duration; a key indicator that, we have a segment of customers whose resources are over-provisioned. There is a need to not only provision new customers moving from on-premise to the cloud correctly, but also right-size existing customers to benefit from cost-saving opportunities. 

\subsubsection*{\textbf{Contribution}}
% As the TAM for migrating on-premise workloads to the cloud is estimated at \$8.1 billion, there is a need for a tool that provides a scalable way to obtain personalized, accurate and secure SKU recommendations. Since no such fully-automated tool \yz{completely replace the manual effort} yet exists today, 
% \sysname offers a scalable and fully-automated approach to obtain personalized, accurate and secure recommendations for cloud providers to map on-premise (and cloud) workloads to appropriate cloud SKUs. 
%\sysname is a data-driven SKU recommendation engine that introduces a novel means of summarizing resource consumption patterns to evaluate the suitability of SKUs and that leverages internal Azure customer data to help guide new customers towards an optimal SKU choice. 
Our contributions are summarized as follows:

\begin{itemize}
	\item Introduce a fully-automated SKU recommendation framework based on \pp curves.
	%circumventing the needs of accessing customer data or queries. 
% 	The method can be generalized to other cloud target offerings.
% 	\item Introduce \pp curves as a way to summarize resource consumption patterns in an easily-understandable way. It can be easily extended to accommodate additional performance dimensions (outside of the current feature set).
% 	and evaluate the suitability of various SKUs for processing customer workloads.\yz{a bit repeating, more concise}
    \item Profile existing Azure SQL customers to understand negotiability around various performance dimensions and preferences with respect to cost/performance trade-offs.
	%\item Provide a customized solution by profiling Azure SQL customers to learn the tolerance levels of resource throttling in each resource dimension, capturing preferences with respect to cost/performance trade-offs. 
	%negotiable resource allocations thus understand their preferences with respect to \pp trade-offs, and leverage the learning to new customers to capture heterogeneity.
	      %\item Produces a personalized recommendation based on customer’s telemetry that accurately identifies the most relevant Azure SQL PaaS.
	      %	\item Provides a straighforward, transparent means for customers to understand the reasoning behind each SKU recommendation in contrast to previous black-box ML strategies. Over- and under-provisioning cases are significantly reduced.
% 	\item Evaluate how \sysname compares to current expensive manual selection methods and an alternative baseline algorithm. 
	%, using extensive telemetry from existing customers in the cloud.
	      %	Evaluated using telemetry from prospective customers and existing customers in the cloud, detect accurately the optimal SKU that matches well with the observed choices and reveal >10\% over-provisioned customers.
	\item Deploy a recommendation engine in the Azure Data Migration Assistant v5.5~\cite{dma,video} that has been used by 600+ unique customers. 
	%has saved \ww{more than X hours} manual assessment for over \ww{y} customers daily. \jc{add up unique instnaces}
	      %	\item Settles the issue from the previous data science approach in that the choice of representative workloads is not required as each recommendation (model) is personalized to that customer.
	      %\item Introduces perf data simulator that allows for the ability to test our strategy, as well as any future SKU recommendation models, in ways that are not currently feasible. 
	      %\item Our synthetic perf data would essentially provide infinite training samples that can be leveraged by many individuals in this migration space to aid in developing ML models that are compatible with the very heterogeneous hardware that exists on-prem and heterogeneous workloads. 
	      %\item We developed this synthetic data to help us debug the console app faster, but such synthetic data should also aid in developing models with greater accuracy and alignment and relevance to real-world behavior, or better tailor their models for specific classes of SQL server workloads / customers (?). Essentially, we are no longer constrained by lack of data. 
\end{itemize}
The remainder of this paper is organized as follows.
%provides a deeper overview of the key challenges of model-driven development in cloud migration and related work.
Section~\ref{sec:related} reviews related work.
Section~\ref{sec:overview} details the offerings in Azure SQL PaaS% and historical migration tools
, and Section~\ref{sec:opt} presents the end-to-end architecture of \sysname.
Section~\ref{sec:integration} discusses production integration efforts.
Section~\ref{sec:result} presents experimental results from back-testing on migrated customers and compares the recommendations from \sysname with an alternative baseline solution.
%Section~\ref{sec:motivation} reviews related work.
%Section~\ref{sec:related} validates our SKU recommendation engine on synthetic workloads.
Remaining challenges and promising directions for future research are stated in Section~\ref{sec:conclude}.

% Despite the utility of these strategies, they require significant input from the customer, and the final recommendations may still be inappropriate. 

% Cloud providers thus default to manual SKU selection as the de facto standard as the decision support systems proposed (e.g.,~\cite{kang2010cloudle, beserra2012cloudstep, kopaneli2015model}) are too hard to use and are difficult to scale.
% Our field experiences have shown that proper SKU selection remains a problem as of 125k databases assessed per year, only 60k proceed with the SKU selections and migrate to Azure SQL.
% %alone accounts for more than \ww{TBD}\%, or an average of \ww{ can take up to 4 weeks}, in the total migration journey \ww{which can take months to finish}.  
% There does not yet exist a straightforward way for customers to exit their existing data centers and land on an optimal cloud SKU~\cite{gholami2016cloud,girish2014survey,alkhalil2017decision}. 
% As a result, a large segment of customers with SQL estates on-premises abstain from moving their workloads to the cloud because of the risks that come with the migration process~\citep{andrikopoulos2014design, andrikopoulos2013supporting, klems2008clouds,gholami2016cloud,girish2014survey,alkhalil2017decision}.
% which includes the possible lack of business continuity, the degradation in performance, and the increase cost in operation. 
% These issues arise because there is no set recipe for migrating customers to the cloud.

%!TEX root = main.tex
%
%\section{Problem Description}\label{sec:background}
%

\section{Background}\label{sec:overview}
%\cmt{General comment, is the section title more suitable as ``Related Work''}

%These cloud services include SaaS (software as a service), PaaS (platform as a service), and IaaS (infrastructure as a service)---which are all accessible and releasable on-the-fly and benefit from the pay-as-you-go model.
%Given this wide landscape of cloud offerings and the fact that there are a wide range of migration tools, from ones that target small- to medium-sized enterprises \citep{wilson2016towards} to those that are specifically designed for scientific applications \citep{saez2015performance}, and that 
%multiple decision points that influence the migration process (e.g., cost, performance, security, interoperability, vendor lock-in, etc.~\citep{wilson2016towards,saez2015performance}) and
As there is a wide range of cloud offerings, e.g., SaaS (software as a service), PaaS (platform as a service), and IaaS (infrastructure as a service),
we narrow the scope of this study to focus on migrating on-premise SQL workloads to Azure SQL PaaS solutions, which includes \textit{Azure SQL Database (DB)} and \textit{Azure SQL Managed Instance (MI)} \citep{SQLMI,SQLDB}. Work is ongoing to generalize the \sysname framework to support other migration scenarios, across other database systems like Oracle~\cite{oracle} and PostgreSQL~\cite{postgre}.
% as they are the most preferred migration target platform both for customer as well as recommended by Azure~ \citep{migrate_target}. 
\textit{Azure SQL DB} offers single database deployment options that create fully managed, isolated databases; while \textit{Azure SQL MI} offers fully managed SQL servers that host a large number of databases.
Within the \textit{virtual cores (vCore)} purchasing model~\cite{vcore}, two service tiers are offered: \textit{General Purpose (GP)} and \textit{Business Critical (BC)}. The \textit{BC} tier offers higher transaction rates and lower-latency I/O compared to that of the SKUs in the \textit{GP} tier. Azure currently offers over 200 different PaaS cloud SKUs.
%In addition, we only consider Gen5 generation for hardware as they are the default deployment option.

%From our field experiences, focus groups, and the research literature, 
%While we have had many smooth migration experiences in that customers' were content with their SKU selection; we have also had many difficult experiences due to customers either getting under- or over-provisioned with their on-cloud SKU. 
%With the extensive collection of Azure SQL DB and MI SKUs, customers often fail to select the right SKU for migration, and tend to under- or over-provision. Our study for Azure customers over a 30-day observation window shows that nearly 30\% of SQL databases use only up to 43\% of the provisioned CPU resources. On the other hand, there are 5\% of SQL databases reach the maximum provisioned CPU usage for more than 10\% of the time which indicates under-provision. 
As no tool currently exists that provides personalized, accurate and secure SKU recommendations among this set of 200 cloud targets, the Azure Data Migration Assistant (DMA)~\cite{dma} was conceived to ease and reduce the risks associated with migration. DMA plays a key role in the general Azure Migration Service that provides a unified migration experience for all Azure products.
\begin{figure}[h]
	\centering
% 	\vspace{-0.2cm}
	\includegraphics[width=\columnwidth]{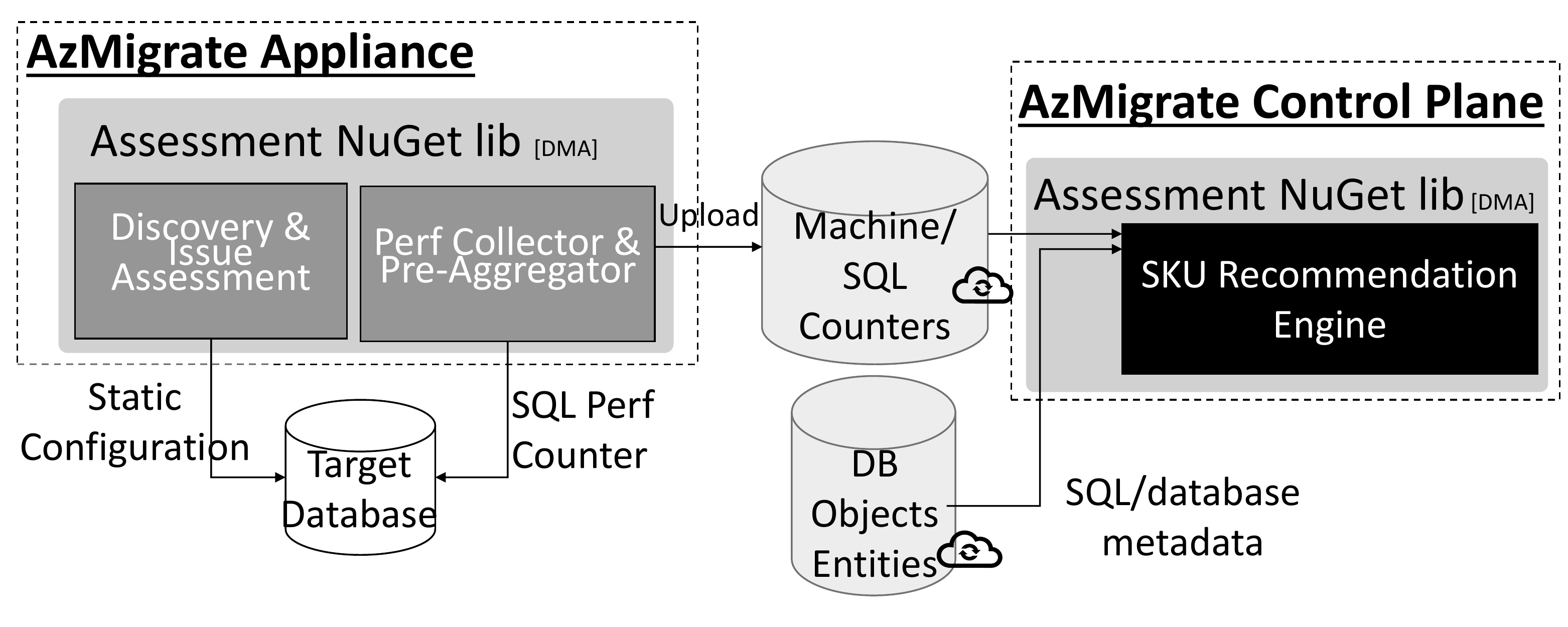}
	\vspace{-0.6cm}
	\caption{How our SKU recommendation engine fits in the current architecture of the Azure Migration Service.}\label{fig:dma}
 	\vspace{-0.2cm}
\end{figure}
%It has an interface directly into the NuGet to enable SKU recommendations for workloads moving to Azure SQL PaaS and facilitate the end-to-end migration assessment. It
\noindent As shown in Figure~\ref{fig:dma}, the AzMigrate Appliance consists of 2 modules: (1) Discovery \& Issue Assessment to detect compatibility issues that would either block or hinder the customer's ability to migrate (e.g., unsupported, or partially supported, features that are currently in use on-premise) and (2) Performance Collector \& Pre-Aggregator to gather SQL performance (perf) counters on CPU, storage, memory, IOPs, and latency. The static configuration and the perf counters are first stored locally on the target database. The perf counters are subsequently uploaded, along with SQL DB metadata, as input to the recommendation module in the Control Plane, where the following two approaches are implemented in the recommendation engine:

\begin{itemize}
	\item \textbf{Baseline strategy.} As \sysname was not available when the DMA tool was first conceived, a naive, baseline algorithm was introduced. This SKU selection procedure involves taking the entire time-series vector collected on each available perf counter (e.g., CPU, memory) and collapsing it into one scalar value. Field engineers often chose either the max of each perf vector, or some large (95\%) quantile. From these values, the cheapest Azure PaaS offering that satisfies all the requirements is suggested. Since max (or large percentiles) determine the customer's resource needs, this approach generally results in over-provisioning.
	      %How the current mapping heuristic fits into the migration process is illustrated in Figure~\ref{fig:dma}. 
	      %\change{This approach has been in production for \ww{6 months} and has led to \ww{123} successful migrations to date. However, \ww{X\%} of customers migrated based on these suggestions have since changed their on-cloud SKU, as this strategy often over-provisions the customer, especially when the assessment period includes atypical workloads that raise the CPU, memory or IOPs utilization for brief periods that are not representative of the customers' usual workload.} \cmt{Hard to get these stats as it was offline offering and I think the astypical workloads could affect our case too. Rewrite some as below}. 
	      %This baseline approach is designed to avoid extreme large compute size recommendation and also provide justifications on the recommended SKU. 
	      %However, it is very challenging to determine the resource usage requirement and properly choose the right percentiles to avoid over- or under-provisioning.
	      %the optimal SKU needs to satisfy each resource usage requirement according to the aggregation methods (quantile or max). 
	      %This method is always too conservative and pushes the recommended SKU to higher end that is too expensive if using maximum aggregation. On the other hand, if a lower percentile aggregation was used, the method would recommend a under-provisioned SKU which are not able to meet the actual performance requirement.
	\item \textbf{\sysname (elastic) strategy.}
	      The motivation for \sysname was to improve upon the existing baseline approach. It is the first model-driven recommendation system that has been integrated into the DMA tool in production.
	      Since its release in DMA v5.5~\cite{dma} in October 2021, our recommendation engine has been widely used by a large number of both external and internal customers (see Table~\ref{tab:dmausage}).
	      %\sysname addresses the following deficiencies of previous model-driven implementations:
	      %\begin{itemize}
	      %	\item Avoid cases in which customers are over-provisioned resources and thus reduce the high cost of migration based on the \pp curve strategy;
	      %	\item Customize to account for heterogeneity between customer preferences as some customers may be more sensitive to certain performance trade-offs in a particular resource dimension; and
	      %	\item Provide SKU recommendations and associated reasoning that is data-driven and easily explainable so customers receive greater transparency on how the recommendations are made.
	      %\end{itemize}
	     In the following sections, we detail this approach and its integration with DMA.
\end{itemize}

%For storage, the Azure Migration Data Plan involves storage on Cosmos DB~\cite{cosmosdb} for SQL and database metadata, and Kusto~\cite{kusto} for SQL and machine requirements and the SQL counters.
%\ww{Include discussion here on long-term plans to embed \sysname into a SKU recommendation service? and discuss Figure~\ref{fig:sku_as_service}.} \ans{Not sure of this part, can we instead introduce the online experience of Azure Migrate and Azure Data Studio as future work? Added somthing below.}

\section{\sysname Engine}\label{sec:opt}

In this section, we present an overview of the \sysname system architecture and the design of each module and its assumptions.
%\change{We assume that customers' are interested in migrating their workloads either to Azure SQL Database or Azure SQL Managed Instance;} \cmt{see red} 

\subsection{Design Principles and Architecture}
The following core design principles capture the motivation behind the architectural design of \sysname.

\mypar{Avoid using customer data/queries} 
With the need to comply with General Data Protection Regulation (GDPR)~\cite{gdpr, gdpr-ca}, and the increasing concerns about data privacy and security, we developed a solution that is solely based on low-level resource consumption statistics. 
% This means using aggregated performance counters as the only input to understand customer workload resource needs. 
% The advantages of this input includes the fact that it is not only less sensitive than customer data/queries, but it can be collected with low overhead. 
From our exploratory efforts, we discovered that we can generally infer various levels of \textit{resource throttling} when we compare customers' performance history against the resource capacities of different SKUs, which is an important input for the recommendation. 
% Our approach allows us to take a much less conservative strategy than the baseline as that assumes zero tolerance towards any type of throttling.

%Instead, we leverage performance counters that are less sensitive and can be collected with low overheads. In face of the increasing concerns about data privacy and security, along with the development , in this project, we develop a solution that is solely based on aggregated performance counters, such as the resource consumption of user workloads, as opposed to the original user query and input data. Combining with the resource capacity for different SKUs to migrate to, we can infer the possible ``performance'' of user workloads given different levels of resource throttling. We set off from exploring simple evaluation of a baseline approach to be very conservative assuming zero tolerance on throttling probability. 

\mypar{Consider cost and marginal benefits} 
As a more expensive SKU results in better performance, 
% As discussed in Section~\ref{sec:intro}, w
we want to match migration customers to the most cost-efficient SKU and avoid over-provisioning, or instances where compute resources sit idle. We introduce our novel \pp methodology as a starting framework for customers to understand the trade-offs between cost and performance (as measured by resource throttling probabilities). 

%The Total Cost of Ownership (TCO) is no-doubly one of the most important factors for customers to select their service. In general, a more expensive product leads to better performance. However, a diminishing return can always be observed if one chooses a very high-end service that always results in idle resources. Therefore, we want to better measure the resource utilization efficiency combining with the price to help customer select the most cost-efficient solution, which leads to the development of the \pp curve approach that can potentially ``sacrifies'' some performance to a lower cost of product.

\mypar{Leverage what we know from existing customers} 
Since we already have a large base of (successfully) migrated customers, we want to learn from their SKU choices.
%\textit{profiling analysis}. 
In particular, we gain some understanding of their level of satisfaction with their migrated SKU by studying their behavior with regards to how often they switched SKUs. Given this telemetry, and their associated performance footprint, we are able to generate their \pp curves and observe where on their respective \pp curves do their SKU choices land. In other words, we have insight into customer preferences for cost over performance, and we leverage this knowledge to cluster customers into various different groups. With these profiles, we can guide new migration customers towards choosing their optimal SKU based on similarities in resource utilization to our existing migrated base.  

%In goal of proposing a more customized solution given different sensitivity to price and the heterogeneity of the workloads (some might be less impacted by resource throttling thus can be more amenable for more economic solutions), we leverage similar ideas of customer segmentation to conclude clustering analysis based on existing Azure customers to better understand their choice of SKUs given different patterns of resource utilization profile and determine the most suitable SKUs for new customers. 

\mypar{Make sure the solution can scale} Since the de facto standard of manual SKU selection takes too long, we need an effective way for new migration customers to quickly get personalized and accurate SKU recommendations. 
Azure Migration Service was designed as highlighted in Figure~\ref{fig:dma} in order to make such automated SKU recommendations possible and support a higher volume of migration.
%with a data collection module that is executed on the customer machine and a recommendation engine that processes the perf data locally. 

%This design allows it to support a larger number of migration efforts. \yz{conflict with Figure 2, whether it is indeed locally or remote, check with Wenjing}
% , and to improve our engine, efforts are currently underway to connect our locally-generated SKU recommendations to centralized storage.

%To support potentially large number of customers' migration, the tool was designed to consist of the data collection module that was executed on customer machine and the recommendation module that process the aggregated data to generate SKU recommendation. The code is well modulized and connected to a centralized storage.

\begin{figure}[t]
	\centering
	%\vspace{-3ex}
	\includegraphics[width=1\columnwidth]{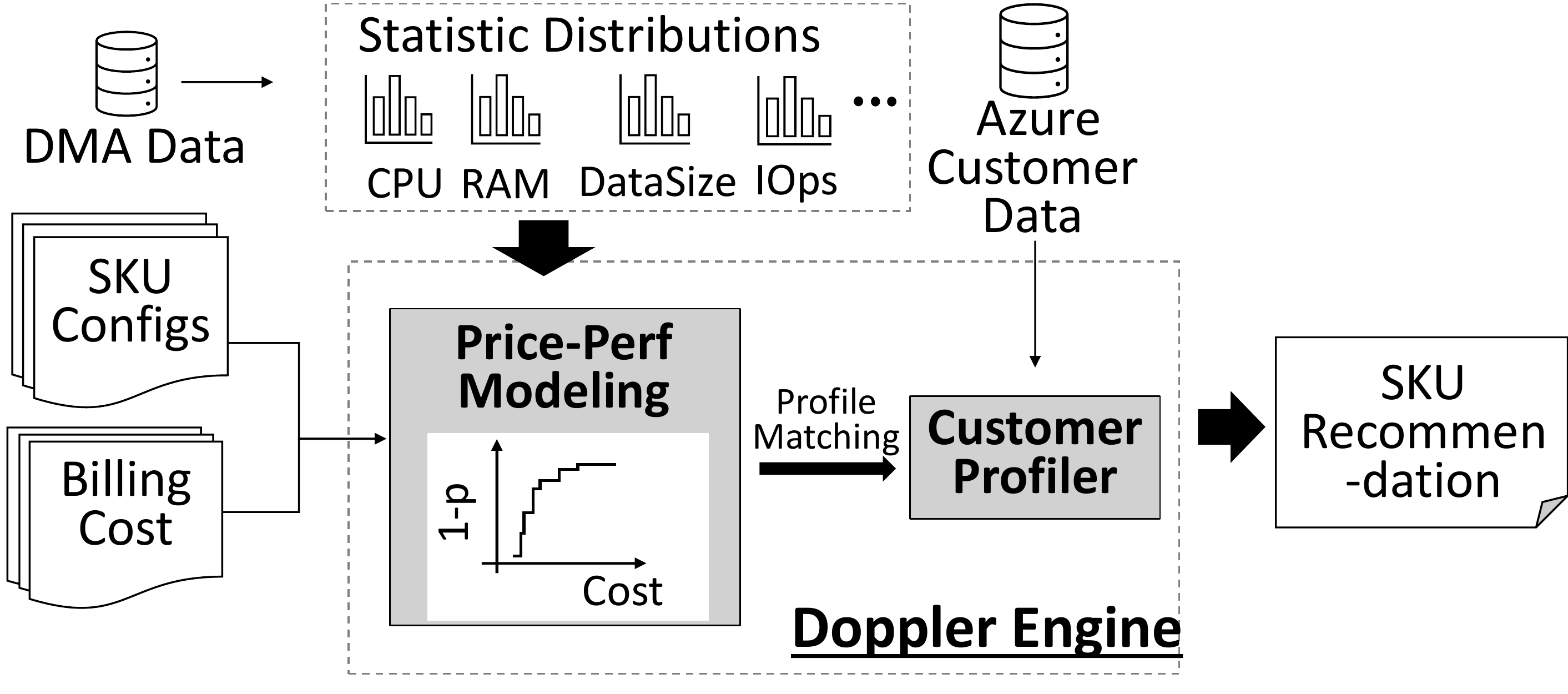}
 \vspace{-0.8cm}
	\caption{The \sysname engine.}\label{fig:doppler}
	\vspace{-0.6cm}
\end{figure}

\sysname achieves the aforementioned design ideas with the system as outlined in Figure~\ref{fig:doppler}. Our SKU recommendation engine consists of two main modules: the \textit{Price-Performance Modeler (PPM)} and the \textit{Customer Profiler}. The \textit{PPM} improves upon the current baseline strategy as it is capable of providing more flexible (elastic) SKU recommendations. It does not require that the SKU meet the customers' (max) resource use, and thus avoids over-provisioning. It takes three inputs: (i) performance counters collected from the DMA tool; (ii) all the possible cloud target PaaS SKUs; and (iii) the real-time pricing associated with each SKU. Since (ii) and (iii) are primarily fixed, the key input to the PPM is the customer performance history, which currently consists of four perf dimensions: CPU, memory, IOPs and latency. By relying on these low-level resource statistics, we circumvent the need to access customer data/queries, addressing privacy concerns. Moreover, unlike the baseline approach that reduces each time-series vector into one scalar value, PPM utilizes the full distribution of each perf dimensions' data. From these inputs, we generate a \pp curve, which effectively provides a personalized rank of all the relevant SKUs a customer can migrate towards.
%makes SKU recommendations more flexible (elastic) by accounting for the full distribution of each performance dimension and avoids the high dimension reduction of the baseline strategy.
With this \pp result and internal insights from customer SKU selections in the cloud, the \textit{Customer Profiler} can match new migration customers to our existing base to help guide them towards one optimal SKU choice. By leveraging the historical decisions from our (successfully) migrated customers, \sysname allows new customers to make a more informed decision.

While the current version of \sysname targets workloads moving to Azure SQL DB and MI %(as they are the preferred target platforms) 
\citep{migrate_target}, \sysname can be easily extended to accommodate additional performance features and adapted to support migration scenarios for different database systems.
%We call this framework in which we make SKU recommendations by first, forming a \pp curve, then second, clustering the customer, the \textit{elastic strategy}. It complements the existing baseline strategy in the DMA tool as the baseline SKU recommendations are effectively based on the max values in a customers' resource profiles. 
%In the following sections, we discuss each module in more detail.

\subsection{Price-Performance Methodology}
\label{sec:explore} 

%\mypar{The price-performance curve}
%The foundation of \sysname lies in generating a personalized price-performance curve.
The \pp curve relates the price, in terms of monthly billing cost, of relevant cloud SKUs to their respective performance, as measured by each SKUs ability to fulfill the customers' resource needs for a pre-specified assessment period. One of the key challenges to \sysname is estimating performance appropriately, especially when it involves different hardware that the workload has not yet been executed upon. Current solutions have averted this challenge by focusing instead on replaying the customer workload on recommended SKUs~\cite{kopaneli2015model}, or if access to customer data is impractical, simulating workloads to replay~\cite{saez2015performance}. By taking these more direct routes, these solutions are able to strictly demonstrate what the performance outcomes will be when migrating workloads to a new SKU. But these strategies also rely on strong assumptions (e.g.,~\cite{leis2021towards}) and extensive data access (e.g.,~\cite{sen2021autoexecutor,zhu2021kea}). We circumvent these problems, as well as, the fact that replay is often not feasible given the large number of possible SKUs, by developing a proxy for estimating performance, which we call the probability of resource throttling.

\begin{figure}
\centering
\begin{subfigure}{.48\columnwidth}
  \centering
  \includegraphics[width=.9\linewidth]{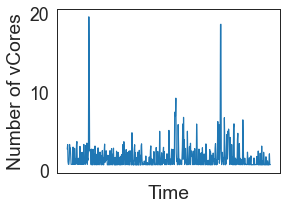}
  \caption{CPU usage by time.}
  \label{fig:ex_weight1}
  \vspace{-0.3cm}
\end{subfigure}%
\begin{subfigure}{.50\columnwidth}
  \centering
  \includegraphics[width=\linewidth]{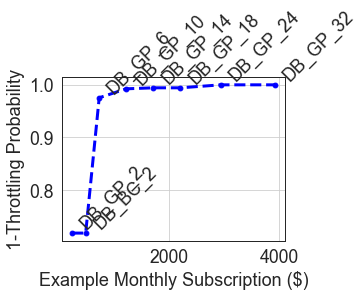}
  \caption{Price-performance curve.}
  \label{fig:ex_weight2}
  \vspace{-0.3cm}
\end{subfigure}
\caption{Example of \pp curve generation from performance history.}
\vspace{-0.4cm}
\label{fig:ex_weight}
\end{figure}

%One of the key challenges of this work is to propose an appropriate way to measure the corresponding performance quantitatively given the what-if scenarios, i.e., if migrating to a particular SKU. Previous work has been focusing on developing a simulation platform~\cite{saez2015performance} or actually replaying customer workloads on all the SKUs~\cite{kopaneli2015model}, which is not feasible given the large number of SKUs and the inaccessibility to customer workload data.Performance prediction of execution time with different hardware is a well-known challenging problem. Existing solutions either rely on strong assumptions (e.g.,~\cite{leis2021towards}) or require extensive data such as query logs (e.g.,~\cite{sen2021autoexecutor,zhu2021kea}). In this research, we propose to measure the probability of resource throttling instead, which provides a good approximation of the customer-oriented performance metrics.
%In other words, the \pp curve provides a personalized ranking of all the relevant cloud SKUs for each customer to migrate towards.
%\red{We also illustrates that \sysname can be easily extended to both Azure SQL DB and Azure SQL MI recommendation in this section.}.
%\yz{a bit jump, forward reference to later replay, believe is a bit too weak}
In Section~\ref{sec:syn}, we demonstrate how this new metric provides a good approximation of the true throttling that a customer might experience, and thus is a sufficient means to approximate how well each SKU meets customer workload performance needs. The \textit{throttling probability} of $\text{SKU}_i$, where $i=1, \ldots, m$ represents some SKU among the $m$ relevant Azure SKUs, is symbolized by $P_n(\text{SKU}_i)$. It is defined as the probability of running into resource throttling in \textit{any} resource dimensions for customer $n$:
\begin{equation}\label{eq:pp}
	\scriptsize
	P_n(\text{SKU}_i) = P(r_{\text{CPU}_n} > R_{\text{CPU}_i} \cup r_{\text{RAM}_n} > R_{\text{RAM}_i} \cup \ldots \cup r_{\text{IOPS}_n} > R_{\text{IOPS}_i}).
	% P(\text{SKU}_i) = P(r_{\text{CPU}} < R_{\text{CPU}_i} \cup r_{\text{RAM}} < R_{\text{RAM}_i} \cup \ldots \cup r_{\text{IOPS}} < R_{\text{IOPS}_i})
\end{equation}
$r_{\{\text{CPU}_n,\text{RAM}_n,...,\text{IOPS}_n\}}$ denotes the vector of random variables corresponding to the resource usage for customer $n$ as collected by the DMA tool, and $R_{\{\text{CPU}_i,\text{RAM}_i,...,\text{IOPS}_i\}}$ denotes the maximum capacity for each perf dimension as fixed by a specific $\text{SKU}_i$. As data becomes available for additional resource dimensions (e.g., wait stats), the throttling probability definition can be extended accordingly. Based on our mathematical formulation above, there are certain performance dimensions that require small adjustments; for example, IO latency is taken as the \textit{inverse} of the actual IO latency in order to calculate the effect of this performance dimension relative to an upper bound ($R_{\text{IOPS}_i}$) for each possible SKU. Figure~\ref{fig:ex_weight} illustrates an example of a customer workload's CPU usage by time and the corresponding throttling probability when only the perf dimension of CPU is factored into this new metric.

%~\footnote{Notice for the IO latency dimension, as each SKU has a lower bound as oppose to an upper bound, we use the inverse of IO Latency in actual creation of \pp curves. This is intepreted as the maximum IOPS that allowed given latency increase in 1 milliseconds. 
%For example the minimum latency of GP tier is 5ms and 1ms for BC tier, then the maximum allowed IOPS given an increased latency in 1ms would be 200 for GP and 1000 for BC.}

%Our proposed elastic strategy is more flexible in that the max values in the resource profiles are considered in the SKU recommendation, but holds equal weight with other values in the customers' resource profile. 

%If a customer tends to exhibit consistent CPU utilization around its average, this average value will be attributed greater weight in the calculation of the \pp curve as highlighted by Figure~\ref{fig:ex_weight}.

%\begin{figure}[t]
%	\centering
%	\includegraphics[width=1\columnwidth]{images/db_resource_usage_corr}
%	\caption{The Correlation Heatmap of Resource Usage}
%	\label{fig:corr}
%\end{figure}

For our SKU recommendation engine, we focus primarily on the four performance dimensions of CPU, memory, IOPs and latency. For customers that are specifically interested in migrating towards Azure SQL DB, we include two additional dimensions of log rate and storage. Since estimating the probability of throttling requires modeling all these resource dimensions \textit{jointly}, we initially considered various %parametric and nonparametric 
statistical methods, such as multivariate kernel density estimation based on vine copulas~\cite{nagler2016evading} and Gaussian smoothing~\cite{silverman2018density}.
%fitting a joint distribution to some known distribution family. %As highlighted in Figure~\ref{fig:joint},
While these existing approaches can do a sufficient job generating the \pp curve, the time it takes to do so is impractical. Given the complexity of these traditional estimation methods, we default to a \textit{non-parametric multi-variate} approach. This estimation technique simply requires calculating the frequency with which all performance dimensions $r_{\{\text{CPU}_n,\text{RAM}_n,...,\text{IOPS}_n\}}$ are satisfied by each SKU, at each time point. Section~\ref{sec:result} highlights how this simple approach can achieve accurate SKU recommendations. For both the parametric and non-parametric approaches considered, we enforce monotonicity on the \pp output so that customers cannot select SKUs that are more expensive and less performant.

\subsubsection*{\textbf{Determining file storage tier for MI}}

\begin{table}[t]
	\caption{File IO characteristics associated with various Azure SQL MI General Purpose (GP) SKUs. \cite{MIResourceLimits}}
	\label{tab:filetier}
 	\vspace{-2mm}
	\resizebox{\columnwidth}{!}{%
		\begin{tabular}{cccccl}
			\toprule
			Storage Tier  & P10            & P20              & ... & P50          & P60          \\
			\midrule
			File size  & $[0, 128]$ GiB & $(128, 512]$ GiB & ... & $(2, 4]$ TiB & $(4, 8]$ TiB \\
			IOPS       & 500            & 2300             & ... & 7500         & 12500        \\
			Throughput & 100 MiB/s      & 150 MiB/s        & ... & 250 MiB/s    & 480 MiB/s    \\
			\bottomrule
		\end{tabular}
		%		\begin{tabular}{ccccccl}
		%	\toprule
		%	File Tier  & P10            & P20              & P30            & P40          & P50          & P60          \\
		%	\midrule
		%	File size  & $[0, 128]$ GiB & $(128, 512]$ GiB & $(0.5, 1]$ TiB & $(1, 2]$ TiB & $(2, 4]$ TiB & $(4, 8]$ TiB \\
		%	IOPS       & 500            & 2300             & 5000           & 7500         & 7500         & 12500        \\
		%	Throughput & 100 MiB/s      & 150 MiB/s        & 200 MiB/s      & 250 MiB/s    & 250 MiB/s    & 480 MiB/s    \\
		%	\bottomrule
		%\end{tabular}
	}
 	\vspace{-4mm}
\end{table}
%\red{
%	Before we move on to the next step of optimal SKU selection, we would like to introduce the necessary changes in order to enable Azure SQL MI recommendation.
%	Azure SQL MI has dynamic resource limits that depend on the file size of each database. Unlike Azure SQL DB from which the resource limits is fixed for each SKU, 
%The recommendation for Azure SQL MI products is different from Azure SQL DB as customers can get dynamic IOPS resource limit by increasing the file size therefore we don't have a fixed IOPS limit in the GP tier but rather depends on the specific data file layout chosen.
%where the \pp modeling cannot be applied directly as discussed. 
%For the General Purpose (GP) tier of Azure SQL MI, unlike standalone Azure SQL DB, every database file gets delicated IOPS and throughput that depends on its file size. This essentially means that customer can get dynamic IOPS resource limit by increasing the file size therefore we don't have a fixed IOPS limit in GP tier but rather depends on the specific data file layout. 
%Table \ref{tab:filetier} shows the current resource limit of different file tiers. In order to utilize the \pp technique to compare all the SKUs,
%	it is essential for us to first decide the IOPS limit for the GP SKU based off of optimal file tier recommendation for each data file. In general, 
%we proposed a two-step \pp curve creation procedure for Azure SQL MI:

In order to make appropriate Azure SQL MI SKU recommendations, we introduce slight adjustments to our resource throttling probability calculation. This is required because, unlike SQL DB SKUs, SQL MI General Purpose (GP) IOPs resource limits are dynamically scaled to accommodate the allocated file size. This happens since the data layer for SQL MI is implemented using Azure Premium Disk storage, and every database file is placed on a separate disk. Each disk has a fixed size, and bigger disks are associated with better throughput and IOPs. From a migration standpoint, since the SKU choice for MI customers begins with fixing the file layout (i.e., a customer can choose an MI SKU that creates 3 files that can each fit within a 128GB disk), we adhere to a similar procedure prior to generating the \pp curve as the IOPs limit $R_{\{\text{IOPs}_i\}}$ is not a fixed variable.
%The IOPs limit $R_{\{\text{IOPs}_i\}}$ is therefore not a fixed variable. \ww{As an example, if a customer has a file layout of $3 \times 128$ GB for an instance, i.e. creates 3 files each with 128 GB storage size, then the maximum IOPs per instance will be 1500.} \yz{what is 3 * 128, is it 3 files? check with Wenjing?}Since file layout is often not negotiable in making relevant SQL MI SKU recommendations, we adhere to the following procedure prior to generating the \pp curve: 
\begin{itemize}
	\item \textbf{Step 1: \emph{Find the correct data storage tier based on data size and workload IOPS and throughput needs}}. We surveyed a number of field engineers to develop this first filtration step to identify appropriate SQL MI SKUs. We filter down the set of possible SQL MI SKUs to only those who can satisfy the \textit{storage} requirement of the data file at a minimum of 100\%. We also verify that this subset of SKUs satisfies at least $95\%$\footnote{The rate $95\%$ is chosen based on file layout analysis of current on-cloud Azure SQL MI resources.
		      %		For MI recommendation, 5 dimensions are used in Equation~\ref{eq:pp}, including CPU, Memory, Data storage size, IOPS and IO latency
	      } of the \textit{IOPS} and \textit{file throughput} requirements.
	      %	calculate the percentage of time that each tier's resource limits can satisfy the usage of IOPS and throughput of the file as a score to rank each file tier. The recommended file tier are the smallest tier that satisfy at least $95\%$\footnote{The rate $95\%$ in \textbf{Step1} is determined based on the file layout analysis for the on cloud Azure SQL MI. For MI recommendation, 5 dimensions are used in Equation~\ref{eq:pp}, namely CPU, Memory, Data storage size, IOPS and IO latency to create the \pp curve.} of the usage time. 
	      In the event that this $95\%$ cannot be reached, we further restrict our search of relevant SKUs to Business Critical (BC) ones in \textbf{Step 2}. Table \ref{tab:filetier} highlights some of the resource limits for different storage tiers. 	      %		Also, if the recommended file tier requires additional storage to be added for the file in order to reach a higher tier that can achieve the IO requirement, we will calculate the minimum size needed to be added to the file. 
	      %		This additional storage will be added on the current storage of the instance when we evaluate the GP SKU in \textbf{Step2}.
	\item \textbf{Step 2: \emph{Create instance-level \pp curves}}. As \textbf{Step 1} parses down the set of SQL MI SKUs only to those that are relevant for the workload at hand, we can then generate the \pp curve as previously described. The only change is that the IOPs limit $R_{\{\text{IOPs}_i\}}$ is calculated based on the file layout chosen---as the summation of IOPs limit on all the data files. 
	
	%The instance-level GP SKU IOPS limit is calculated based on the file layout from \textbf{Step 1} as the summation of IOPS limit of all the data files to create the \pp curve.
	      %		Then follows Algorithm~\ref{alg:pp} for the \pp curve creation.
\end{itemize}

\subsubsection*{\textbf{Limitation}}
While the \pp curve is informative, in that it provides a personalized rank for all the relevant SKUs; its final recommendation is not always clear. The \pp output does not yet take into consideration at which point the customer is willing to negotiate in terms of performance and cost (e.g., if the customer has zero-tolerance for any throttling, are they comfortable with the SKU choice that places them where the \pp curve first hits 100\%?). We initially explored various heuristics to guide the customer towards their ``optimal'' SKU choice, this includes:

%Based on the \pp curve, one straightforward way to propose a recommendation for the SKU is to identify the SKU that located at the position of an ``elbow point'', which indicates a relatively low cost and satisfactory performance. Therefore, \ww{we started with different heuristics approach to determine the optimal SKU.

\begin{itemize}
	\item \textit{Largest Performance Increase}: Selecting the SKU that sits after the point in the \pp curve where the difference in the throttling probability is no longer significant. For example, where $P_n(\text{SKU}_i)-P_n(\text{SKU}_{i-1}) \leq \epsilon$, where $\epsilon$ is set as .001.
	\item \textit{Largest Slope}: Selecting the SKU that sits after the point in the \pp curve that has the largest slope (or rate of change) in the throttling probability. In other words, $\text{SKU}_i$ that maximize $\frac{P_n(\text{SKU}_i)-P_n(\text{SKU}_{i-1})}{\text{Price}(\text{SKU}_i)-\text{Price}(\text{SKU}_{i-1})}$.
	\item \textit{Performance Threshold}: We pick the first SKU whose throttling probability is greater than some predefined threshold, where $P_n(\text{SKU}_i) \geq \gamma$.
\end{itemize}

\begin{figure}
	\centering
	\includegraphics[width=.83\columnwidth]{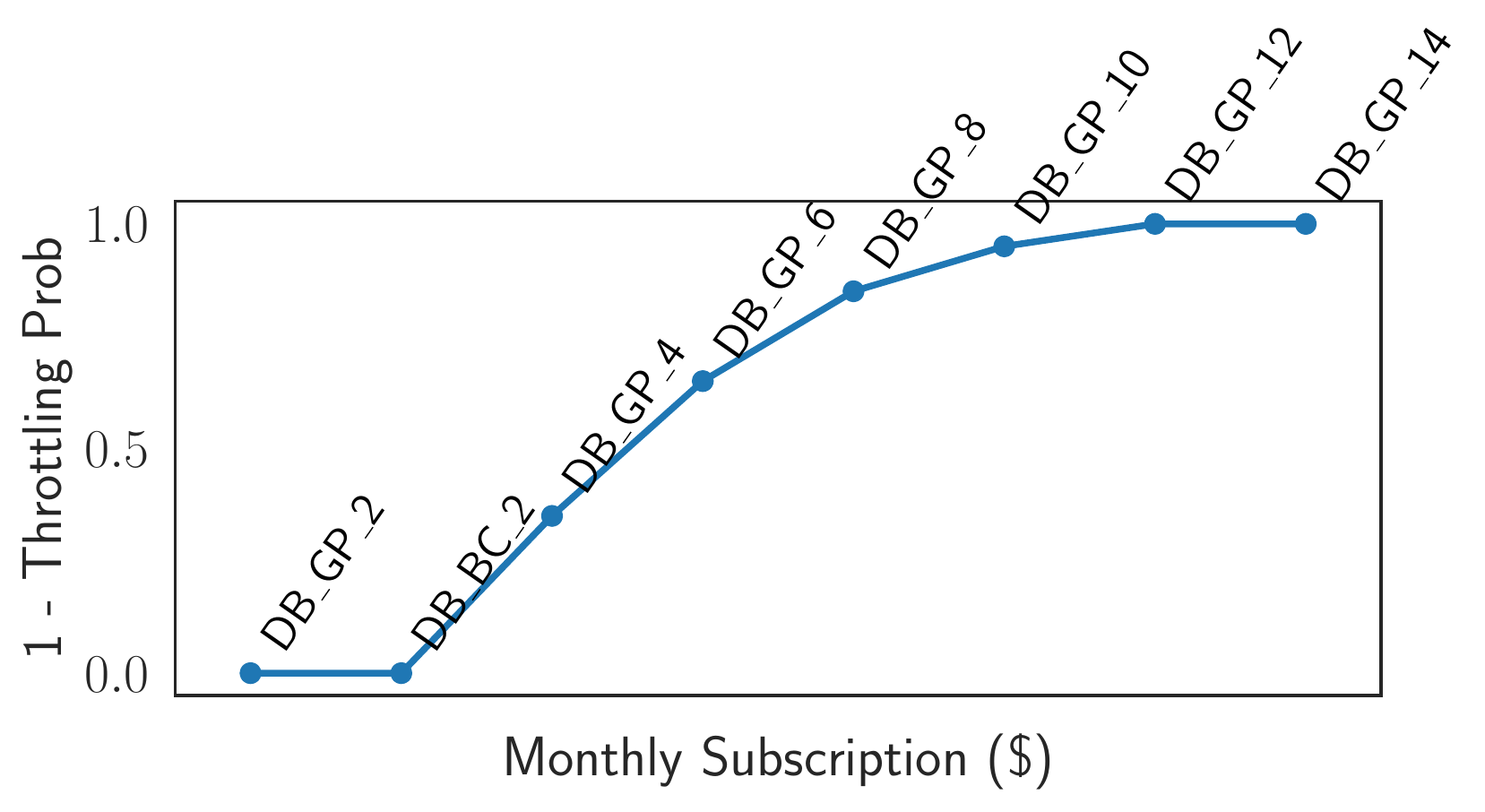}
	\vspace{-0.4cm}
	%\caption{Empirical CDF (ECDF) plot.}
	\caption{Example of a complex \pp curve. Customer chosen SKU is SQL DB General Purpose 14 cores. %\yz{highlight the actually chosen one}
	}\label{fig:elbow}
	\vspace{-.4cm}
\end{figure}

The problem, however, with these heuristics is that they do not always result in the optimal SKU choice. We verify this based on an assessment of existing migrated Azure customers. Assuming that the SKU choice that these customers have fixed (for at least 40 days) is the optimal SKU, we generate the \pp curves for each customer and identify where their fixed SKU choice lands on their respective \pp curves. Neither of these heuristics is able to accurately capture the optimal SKU. Moreover, in the scenarios where the \pp curve is slightly more complex, as shown in Figure~\ref{fig:elbow}, these heuristics are not robust. Namely, following the largest performance increase strategy, this \pp curve would recommend the optimal SKU as SQL DB General Purpose 6 cores; following the largest slope strategy, the optimal SKU would be SQL DB General Purpose 4 cores; and following the performance threshold strategy, the optimal SKU would be General Purpose 12 cores (if $\gamma$ was set to 95\%). 

We also need to consider the limitation that customer workload requirements are highly variable. For example, some applications have high memory requirements; if it runs out of memory, the job will fail, which is worse than running into IO throttling. Other applications might require operating with very low latency, hence delays as mentioned prior, may not be acceptable. Much of the manual SKU recommendation process involves capturing these nuances in workload requirements and understanding what the customer is comfortable with. As this process of measuring workload perf requirements, however, remains much more of an art, we introduced \sysname as one means to better quantify these workload needs and measure what particular performance dimension may be negotiable or not. More specifically, as discussed in the section to follow, we want to capture customers' tolerance for throttling across various types of workloads.
%On the other hand, throttling in some resource dimensions can be more impactful. For instance, if an application runs out of memory, the job would fail, which is more disastrous than running into IO throttling that results in delays. Therefore, a large amount of work for the manual SKU recommendation process was to find the best treatment for each resource dimension respectively, and experienced engineers usually have some sense about what to lookout for. When capturing the resource throttling probability as in Equation~\ref{eq:pp}, different dimensions are treated ``equally'' based on the probabilistic distributions. Though, there are potential to distinguish between case to case based on the distribution ``shape'' of each dimension thus to have different tolerance levels for the throttling probability for different customers/workloads. 

%To , 
We address these limitations and develop a robust means of SKU selection: one that is not as easily influenced by the shape of the \pp curve. Our final strategy, as outlined in the section to follow, leverages our experience with thousands of successfully migrated customers and allows new customers to learn from their choices in a more systematic way. Since \pp curves are able to inherently quantify nuances in each customer's preference, and we are able to study \pp curves generated across Azure SQL PaaS customers at scale, we augment the \sysname framework with a profiling module that clusters customers based on their workload performance behavior and how much they are willing to negotiate on various perf dimensions. By doing so, we provide an explicit means for new customers to see how their own workloads compare, and thus make an informed SKU decision. 
%with a more customized solution that embeds the valuable knowledge on the negotiability of different resource dimensions. 

%\ww{However, we found these heuristics can result in arbitrary SKU recommendation results especially when customer has a complicated price performance curve shape. As shown in example of Figure~\ref{fig:elbow}, the recommended SKU would be GP 6 cores by using largest performance increase approach, GP 4 cores by using largest slope and GP 12 cores by using performance threshold. This can result in potentially under-provisioning as GP 6 cores only satisfy 63\% and over-provisioning as 12 cores satisfy 100\%. Therefore to recommend the optimal SKU, we need a more robust selection approach of the ``elbow point'' against the influence of complicated price performance curve shape.} What's more, the above methods cannot take into account the different preference over prices versus performance for different groups of customers therefore the recommendation was not accurate in terms of matching the existing customer's selection, which gives birth to our method of conducting a comprehensive customer profiling analysis and develop segments of customers to ``learn'' about their preference and apply to the SKU recommendation process. 

\subsection{Customer Profiling}\label{sec:profiler}
Since the \pp curve alone only provides a personalized rank of relevant SKUs for handling new workloads, multiple strategies were tested to develop a principled approach towards selecting one optimal SKU. Unfortunately, the approaches discussed prior that focus purely on the shape of the \pp curve are insufficient for identifying the best SKU. Moreover, these heuristics fail to capture differences in customer preferences for price versus performance as well as sensitivity to resource throttling in different resource dimensions. We thus study resource utilization patterns in each resource dimension, and the respective \pp curves, among customers that have successfully migrated to further distinguish between different types of workloads that might have different tolerance levels for resource throttling. Given customers that exhibit similar resource utilization patterns in the cloud, new customers gain insight into what their optimal SKU might be.

We execute standard ML clustering techniques to profile successfully migrated customers into distinct groups. Extensive interviews with a few of our solution architects provided the domain expertise needed to understand customer preferences in price and performance, and how it manifests in their resource utilization patterns. We characterized customers based on what performance dimension (e.g., CPU, memory, IOPs, latency, etc.) they are willing to ``negotiate'' on. While we did not have the resources to survey each customer, we employed the same rule our customer-facing engineers used: if the \textit{spikiness} of customers' performance counters is rare and short-lived, consider that performance dimension negotiable (for a short period of time). The customer profiling module of \sysname thus characterizes each performance dimension as ``negotiable'' or ``non-negotiable'' by assessing the duration of spikes in the raw time-series (perf counter) data. If the duration is short, the performance dimension is considered less significant, and something that can be negotiated (via cost savings). Based on what performance dimension each customer is willing to negotiate on, they were then clustered into their own respective customer group. 

The calculation for spike duration is straightforward. \sysname first identifies the max peak value(s) within the time-series data of each performance dimension. The variances of the counters are also captured, and a window is formed (one standard deviation) below the max value. The total duration in which resource utilization is within this window is then assessed. If the total duration lasts for greater than a threshold percentage ($\rho$) of the total assessment period, the performance dimension is cast as non-negotiable. Sensitivity analyses were conducted to better tune the $\rho$ threshold. This is referred to as the \textit{threshold algorithm}. Several other techniques were attempted to compare against this thresholding approach; and while some proved useful in summarizing the resource utilization patterns as our field engineers would, we elected this simple thresholding procedure for its transparent interpretation and high performance. Section~\ref{sec:oncloud} compares these competing procedures. %include: \jc{add forward refernece}

\begin{figure}
	\centering
	\begin{subfigure}[b]{0.6\columnwidth}
		\includegraphics[width=\columnwidth]{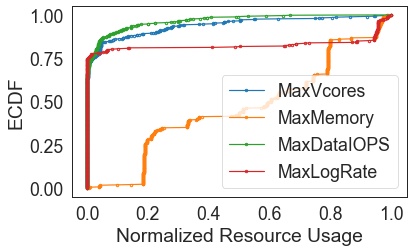}
		\caption{Empirical CDF (ECDF) plots.}
	\end{subfigure}
	
	\begin{subfigure}[b]{\columnwidth}
	    \includegraphics[width=\columnwidth]{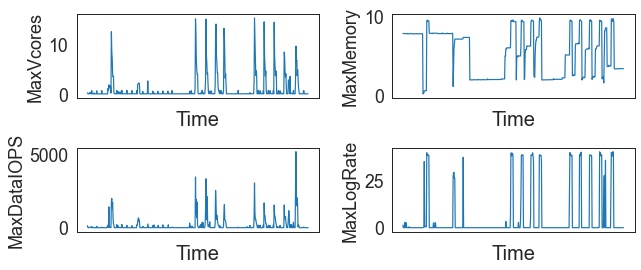}
	\vspace{-0.3cm}
		\caption{Raw time series.}
	\end{subfigure}
% 	\vspace{-5mm}
	\caption{The ECDF and time series associated with various performance dimensions.
		%		a higher ``AUC'' corresponds to more short spiky usage for CPU and IOPS
		}
	\label{fig:auc}
 	\vspace{-5mm}
\end{figure}

\begin{itemize}
	\item \textit{MinMax Scaler AUC}:
	The area under the curve (AUC) is calculated on the empirical cumulative distribution function (ECDF) for each performance dimension. AUC is used as a means to approximate whether the resource is used steadily. As highlighted in Figure~\ref{fig:auc}, higher AUC values tend to describe workloads that had transient spiky usage. For this strategy, the AUC is derived after the performance values are normalized to take values between 0 and 1 (e.g., min-max scaled).

	%Calculated ``area under the curve'' (AUC) metric for each empirical CDF derived from each performance dimension after applying a minmax scaler (normalized to 0-1).
	      %	 i.e $\frac{r_i-min(r_i)}{max(r_i)-min(r_i)}$. \red{
	      %The AUC metric is an approximate description of whether the resource is used more steadily. A higher AUC corresponds to more transient spiky usage (see Figure~\ref{fig:auc}).
	\item \textit{Max Scaler AUC}:
	Similar to the approach listed above, the AUC is calculated after the resource values are only max scaled (e.g., $\frac{r_i}{max(r_i)}$). This calculation better identifies large spikes in resource use.
	      %	\item \textit{Thresholding algorithm}: 
	\item \textit{Outlier percentage}:
	      %	      Detected outliers by examining the deviation from its moving average. 
	      The portion of (performance) counters that exist at least three standard deviations away from the average were calculated as a means to capture spiky usage.
	      
	      %Examine the portion of the samples that have deviation from its moving average exceeding 3 times the full-sample's standard deviation.
	      %	Developed novel shape detection algorithm that detected irregularities in usage in the empirical CDFs. \red{We first interpolate the empirical CDF to create a smoother curve and then fitting a moving average ($m_i$) and standard devation ($d_i$) of lag $X$. We check the spiky usage i.e. $r_i > 3\times d_i + m_i$ and label the resource dimension as spiky if the spiky usage data points are associated with large $d_i$ and are over threshold of $X$.}
	      %	\item \textit{Time series duration}: Assess duration of spikes in the raw time series data. \red{Details for this method is described later.}
	      %	      % \red{We label each time point as spike if it is larger than $max(r_i)-std(r_i)$ where $std(\cdot)$ denotes the standard deviation. If the percentage of the spiky usage is smaller than 1\%, then this usage dimension is labeled as 1 (negotiable).}
	\item \textit{STL variance decomposition}: 
	Using time-series technique, Seasonal \& Trend decomposition using Loess (STL)~\citep{cleveland1990stl}, the observed time-series data ($R_i$) for each perf dimension is decomposed into a trend ($T_i$), seasonality ($S_i$) and residual ($I_i$) component. The residual component is then used for clustering, and transformed as follows: $max(0, 1-\frac{var(I_i)}{var{(R_i)}})$, where $var(\cdot)$ denotes variance. This conversion is meant to capture the variance explained by the trend and seasonality; the closer this value is to 1, the more the observed performance is explained by trend and seasonality.
	%Used popular Seasonal and Trend decomposition using Loess (STL) time-series decomposition \citep{cleveland1990stl} method to decompose the observed usage time series ($R_i$) into trend ($T_i$) $+$ seasonality ($S_i$) $+$ residual ($I_i$). The feature used for clustering is created as $max(0, 1-\frac{var(I_i)}{var{(R_i)}})$  where $var(\cdot)$ denotes variance. This measures the variance explained by trend and seasonality---the closer to 1, the time series are more explainable by trend and seasonality.
	\item \textit{MinMax Scaler AUC result combined with thresholding}: The profiling vector produced from MinMax AUC strategy is concatenated with vectors from the thresholding algorithm to characterize each customer.
\end{itemize}

%\begin{figure}[t]
%	\centering
%	\includegraphics[width=0.95\columnwidth]{example-image-A}
%	\caption{\jc{insert figure here where we increase the 1\% and how results change, need to show why our chosen threshold works best}}\label{fig:tuning_threshold}
%\end{figure}

%Since calculating the time series duration of the max peak values is easily interpretable, and performs with similar accuracy as that of more sophisticated approaches (e.g., STL decomposition), for the version introduced into the next version of DMA (v5.5), this simple heuristic was used. 
To better capture the heterogeneity in tolerance levels of resource throttling in different dimensions, the group membership $g_n$ for a customer $n$ is designed as a function of the negotiability of each resource dimension. In other words, 
\begin{align}\label{eq:membership}
	g_n = f(w_{\text{CPU}_n}, w_{\text{RAM}_n} \ldots w_{\text{IOPS}_n})
\end{align}
where $w_{\{\text{CPU}_n,\text{RAM}_n,...,\text{IOPS}_n\}}$ denotes the negotiability of different resource dimensions as defined in the strategies listed above. A range of standard ML clustering algorithms such as k-means~\cite{hartigan1979algorithm} and hierarchical clustering~\cite{johnson1967hierarchical} can then be executed on the resulting $g_n$ in order to profile customers into different groups.

Based on the grouping of on-cloud customers, we study their preference in price versus performance by examining where their SKU choice lands on their respective \pp curves and examine the corresponding throttling probabilities. 
%preferred degree of the price-performance trade-off by examining the average throttling probability of the chosen SKU in the \pp curve, defined by:
This is formalized as follows:
\begin{equation}\label{eq:Pbar}
	\overline{P}_g = \mathop{\mathbb{E}}_{\forall n,~g_n=g} \left[P_n(SKU_{i_n^*})\right]
\end{equation}
where $SKU_{i_n^*}$ denotes the chosen SKU for customer $n$.
For a new customer $n'$ who is identified to be in the same group, we assume that a SKU with similar  (or slightly lower) throttling probability will be most likely chosen. This SKU can be found by:
\begin{align}
	             & \mathop{\min}_{i}~ \rvert P_{n'}(\text{SKU}_i) - \overline{P}_{g_{n'}} \rvert    \\
	\text{s.t.}~ & g_{n'} = f(w_{\text{CPU}_{n'}}, w_{\text{RAM}_{n'}} \ldots w_{\text{IOPS}_{n'}}) \\
	             & P_{n'}(SKU_i) \leq \overline{P}_{g_{n'}}
\end{align}
\noindent From the competing procedures listed, the thresholding algorithm provides optimal input for clustering as it is able to segment customers into groups whose \pp curves are distinct. One concrete example is shown in Figure~\ref{fig:ex_weight1}, in which there is a segment of customers whose workloads exhibit very short (and uncommon) periods of high CPU utilization. Following the baseline strategy, these customers would be over-provisioned and allocated SKUs that meet this high CPU need. However, following the \sysname framework, these customers would be matched to a cloud user group that avoids selecting the SKU that results in over-provisioning; in fact, for the particular customer whose workload is highlighted in Figure~\ref{fig:ex_weight2}, while the cheapest SKU on their \pp curve (that fulfills 100\% of their resource needs) would push them to select an expensive GP machine with 24 cores, with \sysname, they would be able to leverage historical decisions made by similar customers to learn that it is more optimal to select a cheaper SKU. In short, customers that exhibit similar behavior in regard to spiky CPU usage will tend ``negotiate'' in terms of their CPU resources, and select SKUs with fewer cores, knowing that this choice may result in some level of throttling but larger cost savings over time. By profiling current Azure customers, \sysname provides insight for new migration customers on how similar customers have made their SKU decision and guides them towards a more cost-effective SKU choice.

%Some of the future work can be to extend the classification method to regression models to predict directly the elbow point' throttling probability, however, the accuracy requirement is much higher to ensure the robustness of the recommendation.
%Customers in each group tend to choose SKUs with different throttling probabilities.
%\vspace{-12ex}

%One particular use-case for \sysname is highlighted by the segment of customers that exhibit very short and uncommon periods of high CPU utilization, an example of which is shown in Figure~\ref{fig:ex_weight}. As particular phases of high utilization are not typical and not reflective of the standard workload, with \sysname, we can automatically adjust for these anomalies in the step in which each customer is matched to a cloud user group. In a group with such utilization, customers are more likely to negotiate on that particular resource dimension, thus the picked SKUs tend to have higher throttling probability. In this sense, new migration customers can effectively learn from this SKU decision (with high throttling probability) without making the mistake of over-provisioning themselves. As a result, customers obtain recommendations that are more cost-efficient, in that the recommendation is not only based on their max utilization, but also based on existing users' utilization patterns in the historic data.

\subsection{Confidence Score}\label{sec:confidence}
Since the optimal SKU recommended by \sysname is sensitive to the time window in which performance counters are collected, a secondary metric---the \textit{confidence score}---is developed in order to provide the customer additional certainty towards the final result. This confidence score is derived by bootstrapping the raw customer performance data, generating the respective \pp curve, profiling the workload based on the bootstrapped data, and obtaining the optimal SKU from this process multiple times. By using a random subset of the data, and repeating the process to generate a \sysname recommendation, we get a better idea of how robust the original SKU recommendation may be. The confidence score is the proportion of bootstrapped runs that have the same recommendation as the original.

%evaluates the robustness of the algorithm for the recommended SKU(s) based on bootstrapping of the raw customer profiling data (see Figure~\ref{fig:confidence_score}). Based on the time-series data for the resource usage, we conduct bootstrapping by generating SKU recommendations using a random subset of the original data (e.g., half of the time period, etc.). By repeating the sampling process for multiple runs and deriving the SKU recommendation for each run using the \sysname approach, the confidence score of each SKU can be derived. The fraction of runs that have this recommendation indicates the corresponding confidence level.

\begin{figure}[t]
	\centering
	\includegraphics[width=0.98\columnwidth]{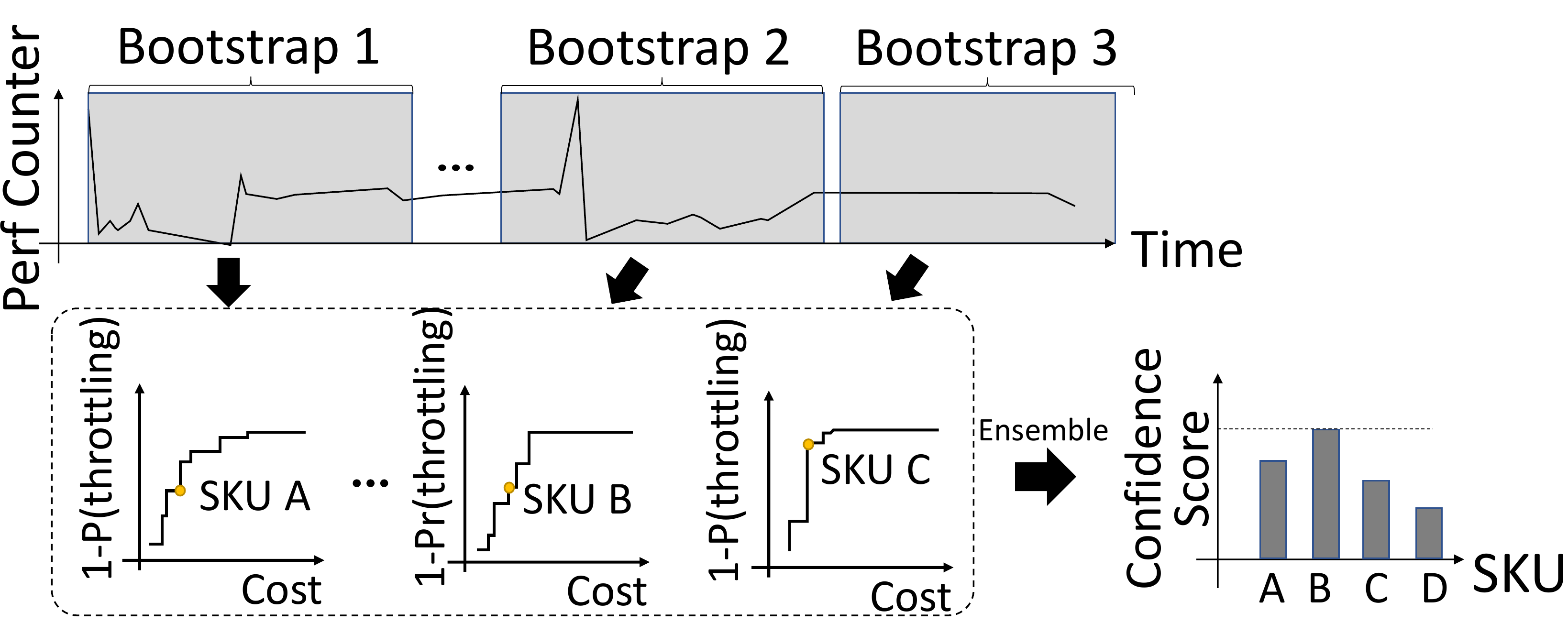}
 	\vspace{-3mm}
	\caption{Confidence score based on bootstrapping samples.
		%		The customer originally fixed the SKU to Azure SQL DB GP vCore 2 on the dotted \pp curve; however retaining this SKU with the new workload demands as highlighted by the solid \pp curve is insufficient. To meet utilization needs at >90\%, the customer changed their SKU to Azure SQL BC vCore 6.
	}\label{fig:confidence_score}
 	\vspace{-5mm}
\end{figure}

As discussed in the section to follow, we typically find that resource utilization patterns that are stable (e.g., no spiky usage) generally have SKU choices made with high confidence scores. %Image that the utilization is a constant, in all the bootstrapping sampled data, the recommendation will be the same. 
In cases where utilization is not consistent, we typically get SKU choices that are associated with low confidence scores. In these particular scenarios, the confidence score can act as a guardrail, suggesting that the customer run the DMA tool for a longer period of time and collect more data (e.g., 1-day's data is often not sufficient to capture standard workload behavior).

%Intuitively, if the resource usage is stable or with repetitive patterns, the recommendation is more consistent as each run of sampling is likely to have a similar usage pattern. This score can be used to determine if we have received a sufficient amount of data for the assessment (sometimes we only receive 1-day's data which might not be representative of the full/complete customer workload). The confidence score is used as a guardrail to determine if we require customers to provide data for a longer period of time.

\section{DMA Integration}
\label{sec:integration}
%\change{Since we do not have any ground truth labels for migration customers,} \cmt{I think we need to integrate to DMA whether we have the labels or not}. 

Given the accuracy at which \sysname is able to pinpoint optimal SKU choices from \pp curves, it has been integrated into the DMA tool~\cite{dma} and publicly released in October 2021.
%, so that new migration customers have access to the SKU recommendations provided via the proposed elastic strategy. 
Three modules were specifically developed for the DMA tool to support the (\sysname) elastic approach:

%\red{
%	During the assessment phase, the raw performance data are collected at a 30 seconds (or specified by customer) interval and aggregated every 10 minutes to save volume of raw data. For every 10 minutes, it reports the maximum, mean, standard deviation, occurrence for different dynamic counters such as CPU, memory at instance/database level, IO at file level. It also collects static counters such as the maximum limits of CPU and memory. These files are saved in a folder for later SKU recommendation purpose. On the SKU recommendation phase, the recommendation engine takes in the aggregated data and conduct further aggregation to find the matching SKU.
%}
\textbf{Data Preprocessing Module} transforms the raw time-series data from perf counters into a format that can be ingested by the \sysname recommendation engine. Given that the existing baseline strategy compresses the original data into one scalar value, this separate module is needed to avoid such high dimension reduction. Since the DMA tool is designed to run for several days, preferably weeks, perf counters are collected every 10 minutes, then aggregated at the file, %\yz{what is file level? check with Wenjing~}, 
database and instance levels. 
%Data are kept in memory. 
Additional inputs of relevant SKU resource limits and customer profiles as illustrated in Figure~\ref{fig:doppler} are calculated offline and saved in the application as static input. A billing interface exists to compute the prices for each SKU. 
	
%	      We extracted maximum value reported every 10 minutes and aggregate into
	      % do the corresponding table pivoting and calculation to aggregate the raw data into 
	      %	      . 
	      %The data is kept in memory. The file recording SKU resource limits (e.g., Figure~\ref{fig:ex_weight}), and preferred scores from the profiler analysis (e.g., Table~\ref{tab:clustering_groups_mi}) are calculated off-line and kept within the application as static input while we utilize the billing interface to calculate the storage and compute the corresponding prices.% for each individual database/instance.
\textbf{SKU Recommendation Pipeline} runs the \sysname Engine to
	      %	is the core component of elastic SKU recommendation engine. The pipeline 
	      build customized \pp curves and recommend the optimal SKU based on customer usage profiling. This pipeline depends on the performance counter input, the customer profiling results and relevant SKUs from the data preprocessing module. 
	      %	It can either recommend SKU based on customer specified target platform (SQLDB, SQLMI etc.), or in the event the target platform is not provided, we compare the total cost of recommended SQLMI and summation cost of SQLDB(s) to output the cheaper choice as the final recommendation.
	      
\textbf{Resource Use Module} provides a visualization dashboard for customers to better understand their workload resource needs. 
	%We also have a separate on-prem workload profiling analysis module to provide more insights for current customer's resource usage. 
	It outputs time series and distribution plots of customer usage across various perf dimensions, as well as, the \pp curve, so that customers can understand why they received a specific SKU recommendation.

%\change{Initial results appear to suggest that customers are content with their DB and MI SKU recommendations. \ww{we may want to include some case studies of customers that use the elastic strategy rec, and track them for a few weeks or a few months to demonstrate that our SKU recommendation was sufficient?}}\cmt{Since this is offline experience, we don't have way to track unless we reach to customers or customers reach us, we could ask PM's input on this.} 
% \yz{can we say: Note that the runtime for all the above modules are on customer's local machine to avoid data transaction and protect data privacy.}
The runtime for all the above modules is installed on customers' local machines to 
%preserve data in-situ in customer machines and 
protect user privacy.
This makes it difficult to automatically track recommendations produced by \sysname as they are currently stored locally. There are surveys that track how content new migration customers are with their \sysname SKU; but efforts are underway to integrate the DMA tool with Azure Migrate Service, which will provide an online means to track every step of a customers' migration journey. We will ultimately be able to keep a record of all the recommended SKUs from \sysname and whether these SKUs were selected for migration, and we will be able to examine the retention of each customer. This feedback loop will be integrated in the \sysname framework, to improve our customer profiling module and help us better understand the preferences of successfully migrated customers. 

\section{Experiments}\label{sec:result}
% \cmt{General comments for this section, since we don't have the large amount of on-premises customer data from Azure Migrate (not sure how long), if we cannot get the data and do evaluation sooner, can we change the main focus of this section to Impact and Future work? Bascially move the summary from Section~\ref{sec:opt} and the part below on future work and avoid summary on comparisons.}

%As \sysname was developed to make SKU recommendations for new migration customers, 
%In this section, 
We evaluate the (\sysname) elastic strategy approach using performance history from customers that ran workloads on Azure SQL PaaS and customers that ran workloads exclusively in on-premise data platforms. For cloud customers, we examined the data from customers that ran workloads on Azure SQL DB and SQL MI between June 2020 and March 2021. We filtered down this set to perf counters collected on 9,295 SQL MI and 7,041 SQL DB as this subset includes customers that have fixed their SKU choice for at least 40 days. In order to test how well \sysname works as a SKU recommendation engine, we assume that these migrated customers that have fixed the SKU choice for this duration are satisfied with their SKU, and thus, their fixed SKU is the optimal choice. This assumption of SKU retention as the ``optimal'' SKU is thus utilized to back-test \sysname. If our framework can match this SKU choice, we assume that we are ``correct'' and that \sysname is an effective engine for mapping on-premise (and cloud) workloads to the optimal cloud target. For customers that have workloads running exclusively on-premises, we study the performance footprint from 257 SQL servers with 1,974 databases collected from Azure Migrate. For such customers, to evaluate the accuracy of the recommended SKUs, we generate synthetic workloads that mimic the performance history from that of the customer, and we execute these workloads on a subset of machines to test the appropriateness of our SKU choice.
%For on-cloud, we collected the performance counters of Azure SQL DB and SQL MI which are active from June 1st, 2020 to March 1st, 2021 that lived longer than 40 days. We are able to select 9295 (80\%) SQL MI and 7041 (91\%) SQL DB that have active usage within the last 14 days and stayed on the same SKU. We made the assumption that those migrated customers are satisfied with the current SKU on account of stable SKU retention. 
%For on-prem data, we tested on the performance counters collected from Azure Migrate in two productions regions for 257 SQL servers with 1974 databases.
%To further evaluate the performance of recommended SKUs, we generate synthetic workload to mimic the ones from customers and execute it on a selection of machines to examine its performance. 

We start our experimental efforts by first comparing the \pp curves for cloud customers versus on-prem in Section~\ref{sec:typical}. We then backtest the \sysname SKU recommendation engine over cloud migration data to understand how well the \pp curves and subsequent customer profiling work in identifying the optimal cloud target in Section~\ref{sec:oncloud}. 
% Confidence scores are also examined in Section~\ref{sec:con-score} along with an exploration of customers that changed SKUs in Section~\ref{sec:changingSKU} to demonstrate how \sysname can be translated to right-size existing customers. 
Since there is no ground truth label for the optimal SKU for new migration customers, running workloads in on-premise data platforms, we benchmark our SKU recommendations against the baseline algorithm in Section~\ref{sec:baseline}. We also leverage synthetic workloads to validate the performance of our recommended SKUs in Section~\ref{sec:syn}.

\subsection{Typical Price-Performance Curves}\label{sec:typical}

\begin{figure}
        \centering
        \begin{subfigure}[b]{0.475\columnwidth}
            \centering
            \includegraphics[width=1\columnwidth]{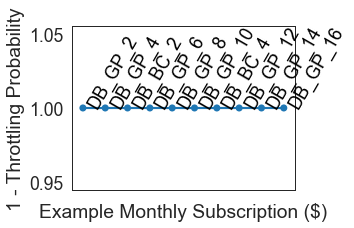}
		\caption{Flat. }\label{fig:ppcurvtype0_simple}
        \end{subfigure}
        \hfill
        \begin{subfigure}[b]{0.475\columnwidth}  
            \centering 
            \includegraphics[width=1\columnwidth]{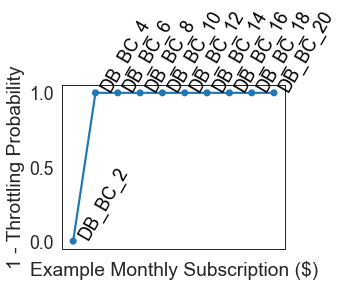}
			\caption{Simple.}\label{fig:simple_pp}
        \end{subfigure}
        \vskip\baselineskip
        \vspace{-0.3cm}
        \begin{subfigure}[b]{0.475\columnwidth}   
            \centering 
            \includegraphics[width=1\columnwidth]{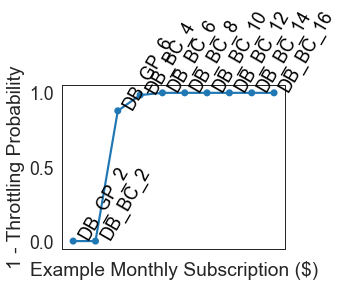}
			\caption{Complex I}\label{fig:odd_pp1}
        \end{subfigure}
        \hfill
        \begin{subfigure}[b]{0.475\columnwidth}   
            \centering 
            \includegraphics[width=1\columnwidth]{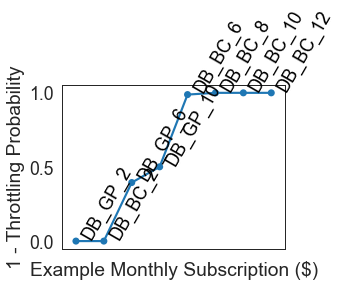}
			\caption{Complex II.}\label{fig:odd_pp2}
        \end{subfigure}
    \vspace{-0.3cms}
       	\caption{Major types of \pp curves.}\label{fig:ppcurvtype0}
    \end{figure}

\begin{figure}[t!]
	\centering
	\vspace{-2mm}
	\includegraphics[width=0.7\linewidth]{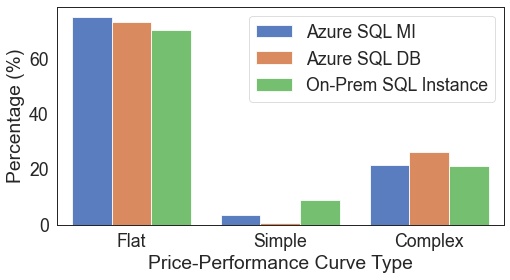}
	\vspace{-3mm}
	\caption{Breakdown of different \pp curve types within our training data set.}
	\vspace{-2mm}
	\label{fig:ppcurvtype}
\end{figure}

There are three typical \pp curve shapes (see Figure~\ref{fig:ppcurvtype0}):
\begin{itemize}
	\item \textbf{Flat}: All relevant SKUs satisfy 100\% of the customers' workload resource needs.
	\item \textbf{Simple}: There is a bifurcation between SKUs that either satisfy 100\% or 0\% of resource needs. Given this division, the cheapest SKU that results in 0\% throttling probability is the clear choice.
	\item \textbf{Complex}: The \pp curve results in a rank of a wide range of SKUs with a variety of throttling probabilities.
\end{itemize}
%flat (Figure~\ref{fig:ppcurvtype0}), simple (Figure~\ref{fig:simple_pp}), and complex (Figures~\ref{fig:odd_pp1} and \ref{fig:odd_pp2}). 
Based on the performance histories of workloads running on Azure SQL MI and DB,
%as illustrated in Figures~\ref{fig:ppcurvtype}, 
73.3\% of this subset of SQL DB customers and 74.9\% of SQL MI customers exhibit a flat \pp curve (see Figure~\ref{fig:ppcurvtype}). In these scenarios, \sysname recommends the cheapest SKU as it is the most cost-efficient option. Among this set, we are able to identify approximately 10\% of customers that were over-provisioned, as their fixed SKU choice places them much farther along their \pp curve. There are a few customers that were paying for SKUs that satisfied 4$\times$ their max resource needs. These particular cases are indicative of customers that are highly over-provisioned; \sysname thus is also useful as a means to identify and right-size existing customers.

On the other hand, 26.2\% of this subset of SQL DB and 21.7\% of SQL MI customers fall in the complex \pp category. While this segment does not appear significant, compared to the proportion of customers with flat \pp curves, they account for a large fraction of Azure SQL PaaS revenue. In fact, this subset of customers accounts for more than \textit{all} Azure SQL PaaS revenue from flat \pp customers combined, hence the need for the \sysname framework. %as a means to choose the cheapest SKU that satisfies 100\% of resource is not satisfactory. 

%\ww{We also discovered that for the nearly 20\% of customer that exhibiting more complex \pp curve shape, they together generate more revenue than the customers shows flat \pp curve shape combined. Therefore, it is crucial to provide good recommendation for both groups. } \yz{can we say the importance of other groups' customers who are more likely to have higher expense in general, which represents large enterprise customers, therefore, it is crucial to provide good recommendation for them as well.} However, we observed that over 10\% of Azure SQL DB and SQL MI customers have chosen significantly higher-end SKUs (sometimes with 4$\times$ larger number of vCores). These flat \pp curves are indicative of Azure customers that are over-provisioned, and the potential monetary saving can be significant.
%For the simple curves, the optimal SKU is clear as there exists a sharp drop in the \pp score from the SKU that satisfies 100\% of the resource needs to other cheaper SKUs. 
%in which the optimal SKU based on the shape of the \pp curve is not clear. For complex curves, 
%where an optimal elbow point with a specific throttling probability needs to be selected based on the customer profiling analysis.
%

%\begin{figure}
	%\centering
	%\begin{subfigure}[b]{\columnwidth}
	%	\centering
	%	\includegraphics[width=\columnwidth]{images/curve_type}
	%\end{subfigure}
	%\caption{Major types of \pp curves.}\label{fig:ppcurvtype0}
%\end{figure}

% \begin{figure}[t]
% 	\centering
% 	\includegraphics[width=0.95\columnwidth]{images/example_flat_pp.png}
% 	\caption{A flat \pp curve. }\label{fig:ppcurvtype0}
% \end{figure}

% \begin{figure}[t]
% 	\centering
% 	\includegraphics[width=0.95\columnwidth]{images/pp_shape_simple.png}
% 	\caption{A simple \pp curve.}\label{fig:simple_pp}
% \end{figure}

% \begin{figure}[t]
% 	\centering
% 	\includegraphics[width=0.95\columnwidth]{images/pp_shape_complex_1.png}
% 	\caption{A complex \pp curve}\label{fig:odd_pp1}
% \end{figure}

% \begin{figure}[t]
% 	\centering
% 	\includegraphics[width=0.95\columnwidth]{images/pp_shape_complex_2.png}
% 	\caption{Another example of complex \pp curve. \yz{ideally the 4 can be merged into one}}\label{fig:odd_pp2}
% \end{figure}

%We also provide anecdotal evidence of how .. can help
% As further discussed in Figures~\ref{fig:ppcurvtype}, \jc{50\%} of \pp curves for Azure SQL DB customers and 56.9\% of \pp curves for Azure SQL MI customers have one clear elbow point, in which the cheapest SKU fulfills the customers' workload requirements at 100\%; however, 26.2\% and 21.7\%, of SQL DB and MI, respectively, fall in the complex category in which the optimal SKU based on the shape of the \pp curve is not clear. \yz{this paragraph is a bit duplicated with the previous one}

% \begin{figure}[t!]
% 	\centering
% 	\begin{subfigure}{.22\textwidth}
% 		\centering
% 		\includegraphics[width=.8\linewidth]{images/db_breakdown_pp_shape.jpeg}
% 		\caption{\pp Curves Type.}
% 	\end{subfigure}
% 	\begin{subfigure}{.22\textwidth}
% 		\centering
% 		\includegraphics[width=.8\linewidth]{images/MI_breakdown_pp_shape.jpeg}
% 		\caption{\pp Curves Type.}
% 	\end{subfigure}
% 	\caption{The percentage breakdown of Azure SQL DB and MI resources assessed that have flat, simple, versus complex \pp curves.}
% 	\label{fig:ppcurvtype}
% \end{figure}
The same breakdown of the three typical \pp curves also exists when applied to performance histories from workloads that are run exclusively in on-premise data platforms (see Figure~\ref{fig:ppcurvtype}).

%the personalized \pp curves look similar to customers already on cloud .
%For those new customers, 70.3\% have flat \pp curves and about 9\% of customers have simple \pp curves.
% in which the best SKU was clear as the SKUs either satisfied the customers' needs at 100\% or satisfied at less than 50\%. 
%In these ``simple'' cases, the cheapest SKU that met the customers' resources at 100\% was recommended. 
%And the rest 21\% of customers have complex \pp curves.
% in which our customer profiling plays an effect, in guiding the customer towards one optimal SKU by learning from the experiences of existing Azure customers.
\subsection{Backtesting with Cloud Data}\label{sec:oncloud}
As there does not yet exist a scalable SKU recommendation engine for selecting optimal SKUs, that map on-premise workloads to the cloud, we need a means to test the accuracy of \sysname. We accomplish this by leveraging internal data we have on successfully migrated customers in Azure and assume that customers that have fixed their cloud SKU for at least 40 days have selected the optimal SKU for their workload needs. We also exclude over-provisioned customers as identified in the previous section. The frequency at which \sysname can match the same (fixed) SKU as these customers is taken as one proxy to measure the utility (accuracy) of \sysname. 
\vspace{-4mm}

\begin{center}
	\begin{table}[t!]
		\caption{Scores associated with each Azure SQL MI customer group (differentiated by the performance dimension negotiability in which 0 denotes negotiable).} \label{tab:clustering_groups_mi}
 		\vspace{-2mm}
		\begin{tabular}{ccccc}
			\toprule
			\textbf{Group} & \textbf{vCores} & \textbf{Memory} & \textbf{IOPS} & \textbf{Average (Std) Score} \\ \midrule
			1              & 0               & 0               & 0             & 0.8500 (0.057)               \\
			2              & 0               & 0               & 1             & 0.9739 (0.054)               \\
			3              & 0               & 1               & 0             & 0.9351 (0.017)               \\
			4              & 0               & 1               & 1             & 0.9692 (0.051)               \\
			5              & 1               & 0               & 0             & 0.9869 (0.026)               \\
			6              & 1               & 0               & 1             & 0.9974 (0.045)               \\
			7              & 1               & 1               & 0             & 0.9668 (0.015)               \\
			8              & 1               & 1               & 1             & 0.9974 (0.056)               \\ \bottomrule
		\end{tabular}
 		\vspace{-2mm}
	\end{table}
\end{center}

\begin{center}
	\begin{table}[t!]
		\caption{Accuracy of \sysname in identifying the optimal SKU based on standard k-means clustering.}  \label{tab:clustering_comparisons}
 		\vspace{-2mm}
		\begin{tabular}{ccc}
			\toprule
			\textbf{Negotiability Definition }         & \textbf{DB} & \textbf{MI} \\ \midrule
			MinMax Scaler AUC                          & 77.3\%      & 74.3\%      \\
			Max Scaler   AUC                           & 78.5\%      & 73.9\%      \\
			\textbf{Thresholding   Algorithm}          & 77.6\%      & 75.1\%      \\
			Outlier percentage                         & 78.1\%      & 74.1\%      \\
			STL  Variance Decomposition                & 78.1\%      & 74.6\%      \\
			MinMax Scaler AUC adjusted with timeseries & 77.8\%      & 75.5\%      \\ \bottomrule
		\end{tabular}
 		\vspace{-2mm}
	\end{table}
\end{center}
%\begin{center}
%	\begin{table}[]
%		\caption{Elastic strategy performance with time series thresholding algorithm} \label{tab:elastic_strategy_performance}
%		\begin{tabular}{ccc}
%			\toprule
%			\textbf{Customer Type} & \textbf{Accuracy} & \textbf{Micro Accuracy} \\ \midrule
%			DB                     & 77.7\%            & GP: 80.7\% / BC: 47.8\% \\ 
%			MI                     & 78.6\%            & GP: 79.9\% / BC: 65.4\% \\ \bottomrule
%		\end{tabular}
%	\end{table}
%\end{center}
%\ww{are there internal efforts for this? maybe add a little more context? \ans{We have SLO auto deflating/inflating for COGS efficiency but it is by customer's choice whether they want to rescale the DB/MI}} \change{Annual savings for customers based on monthly subscription prices are estimated at \$1.26M and \ww{\$XX} for DB and MI customers, respectively.}\cmt{I am not sure about include the statement, since each customer will have different estimated savings and the total cost summation actually is the revenue we will lose.}

\begin{center}
	\begin{table}[]
		\caption{Elastic strategy performance excluding over-provisioned customers.} \label{tab:elastic_strategy_performance_adjsuted}
		\vspace{-2mm}
		\begin{tabular}{ccc}
			\toprule
			\textbf{Customer Type} & \textbf{Accuracy} & \textbf{Micro Accuracy} \\ \midrule
			DB                     & 89.4\%            & GP: 89.0\% / BC: 95.6\% \\
			MI                     & 96.7\%            & GP: 97.6\% / BC: 86.9\% \\ \bottomrule
		\end{tabular}
	\end{table}
 	\vspace{-6mm}
\end{center}

\subsubsection{Customer Profiler}\label{sec:profilerexp}
As outlined in Section~\ref{sec:profiler}, various strategies were tested to summarize the raw time-series performance data for current cloud customers. Each strategy involves compressing the time-series vector for each performance dimension into one scalar value, such that each customers' workload could be described by a simple vector that represents the negotiability of each resource dimension. 
For example, for a customer in which we can access the workload performance traces for the dimensions of CPU, memory, IOPs and latency, under the thresholding strategy, \sysname can reduce this complex time series matrix into one vector. 
An example of what this output vector might look like is $\langle 0, 0, 1, 1\rangle$, where 1 denotes the performance dimension this specific customer is willing to negotiate on (e.g., IOPs and latency). 
% Since the performance footprint for each cloud customer can be succinctly described with simple vectors, typical ML clustering strategies can then be utilized to group customers into different groups. 
By attempting various summarization strategies (e.g., MinMax AUC, thresholding, etc.) for the time series performance traces, and then trying various standard ML clustering strategies on the summarized output, we are able to arrive at a means of clustering existing customers into distinct groups. 

% We ultimately select our unique strategy (thresholding and enumeration) for the customer profiling module as it results in high accuracy for identifying the ``optimal'' SKU and is the most interpretable approach. 

For SQL DB recommendations, the following perf dimensions are summarized: CPU, memory, IOPs and log rate. Since four dimensions are considered, there are thus $2^4=16$ possible customer groups for workloads that suit SQL DB SKUs. For SQL MI recommendations, only CPU, memory and IOPs are summarized; thus, there are only $2^3=8$ possible customer groups for these such workloads. We tested more complex clustering strategies, but found that straightforward enumeration is sufficient in separating customers into distinct groups. 
%In this section, we evaluate the accuracy of SKU recommendation using back-testing on the migrated Azure customers.
%Given the lack of data (e.g., resource profiles) collected from customers with SQL server instances located on-premises, we apply this elastic strategy---of forming the \pp curve and profiling each new customer---on cloud customers instead. 
%By examining the duration of spikes (less than 1\% or not) using the threshold algorithm, for Azure SQL DB customers, we define each resource dimension among vCores, Memory, IOPS and Log Rate as negotiable or not. As opposed to leveraging ML to automatically cluster workloads, we group them based on the combinations of the negotiability on all dimensions. This results in $2^4=16$ groups for SQL DB; and
% as exhibited for one particular training set in Table~\ref{tab:clustering_groups_db}
%$2^3=8$ groups for SQL MI by assessing the dimension of vCores, Memory, and IOPs (see Table~\ref{tab:clustering_groups_mi}).
%. 
%For example, group 3 indicates the memory usage being negotiable while CPU and IOPS are non-negotiable usage.

The \pp curves of customers within each group given the thresholding algorithm
%of these 16, or 8, customers groups for DB and MI (respectively), 
have similar shapes, and the chosen SKUs land around the same throttling probability values as suggested by the small standard deviation values in Table~\ref{tab:clustering_groups_mi}.
% as shown in Figure~\ref{fig:histogram_cloud}. 
%As a result, after identifying the group membership for a new customer, based on his/her \pp curve, we recommend a SKU with the same score (e.g., 1-throttling probability) as the average score associated with the migrated customers in the same group.
%utilize these average values as a mean to guide new customers towards one SKU decision on their unique \pp curves. 
%In other words,
%depending on whichever customer group a new customer might align with in terms of their resource profiles, 
%the optimal SKU is determined as the SKU on their \pp curve that most closely aligns with the average score from the look up table.
As expected, SQL MI customers in group 1 are willing to negotiate on any of the three perf dimensions considered. As a result, they tend to fix their SKU choice on the \pp curves that have a lower score; in other words, they are willing to experience some level of throttling in order to realize cost savings over time. 
Unlike this particular group, other customer groups (e.g., group 8) are not willing to negotiate at all, hence the choice of SKUs that have a high score and are associated with lower throttling.

%It is observed that customers in group 1 who have no negotiable resource dimensions can be aligned with lower non-throttling probability (0.85) than customers in other groups that have at least one negotiable dimensions. Further investigation shows customers in this group are more cost-sensitive therefore tend to choose a lower-end SKU.
%Tables~\ref{tab:clustering_groups_db} or \ref{tab:clustering_groups_mi}. 

Table~\ref{tab:clustering_comparisons} compares the accuracy of SQL DB and SQL MI SKU recommendations using \sysname across the various summarization strategies discussed in Section~\ref{sec:profiler}. The thresholding algorithm reaches comparable performance as the others; while the Max Scaler AUC performs the best as it better captures nuances of the performance distribution. Since calculating the AUC is more time-consuming, and the accuracy of our (simple) thresholding approach is not too far off from the optimal, the thresholding approach was selected for implementation in DMA.
The final strategy deployed in production utilizes the thresholding algorithm, then employs straightforward enumeration to profile customers into various groups. 
% For thresholding, to identify whether customers are willing to negotiate on a specific perf dimension or not, $\rho$ is set to 0.01.
%Due to the latency requirement, we implement the thresholding algorithm in the production pipeline for the DMA tool.
%Table~\ref{tab:elastic_strategy_performance} further breaks down the evaluation to compare the SKU recommendation accuracy for GP and BC, in which this elastic strategy approach is able to match the fixed SKU choice of cloud customers 77.7\% of the time for DB and 78.6\% of the time for MI. The accuracy for GP SKUs are significantly higher than that of BC. The lower accuracy in the BC SKUs are expected since we only look at the differentiation on BC vs GP based on performance counters. However, customer can choose BC tier for reasons like high availability which is not considered in the current recommendation engine. BC tier offers higher availability by having replications of compute and storage. And this is one of the future direction to include more dimensions in the \pp curve.

%Further examination into the cases when 
The majority of cases when \sysname recommends a SKU that does not match the chosen customer SKU %highlights that it pla indicates that the majority of these cases 
involves a flat \pp curve.
% as shown in Figure~\ref{fig:ppcurvtype0}. 
%cases in which this elastic approach mismatches reveals that the resulting \pp curve are often flat, like the one shown in Figure~\ref{fig:ppcurvtype0}, such that every relevant SKU fulfills the customers' utilization needs at 100\%. 
In these scenarios, \sysname recommends the cheapest SKU that fulfills the customers' utilization needs at 100\%. However, since there is a large proportion (>10\%) of customers that are over-provisioned, the most cost-effective option from \sysname does not match the fixed cloud SKU. Figure~\ref{fig:ppcurvtype0_simple} %(top left) \yz{subtitle} 
illustrates one example where the GP 2 cores machine can easily meet the customers' workload resource needs at 100\%, but this customer instead allocated themselves with an 80 core machine. By right-sizing, the customer realized over \$100k in annual savings. As efforts are underway to right-size highly over-provisioned customers, we remove this particular subset of customers from our original backtesting data set. Table~\ref{tab:elastic_strategy_performance_adjsuted} highlights how the accuracy of \sysname drastically improves when over-provisioned customers are excluded from the ground truth labels.

\subsubsection{Confidence Score}\label{sec:con-score}
\begin{figure}[]
	\centering
% 	\begin{subfigure}[b]{0.44\columnwidth}
% 		\includegraphics[width=\columnwidth]{images/confidence_score1_cut.pdf}
% 		\caption{Distribution of the highest confidence score.}
% 		\label{fig:ConfidenceScoreDist}
% 	\end{subfigure}~
% 	\begin{subfigure}[b]{0.545\columnwidth}
		\includegraphics[width=.56\columnwidth]{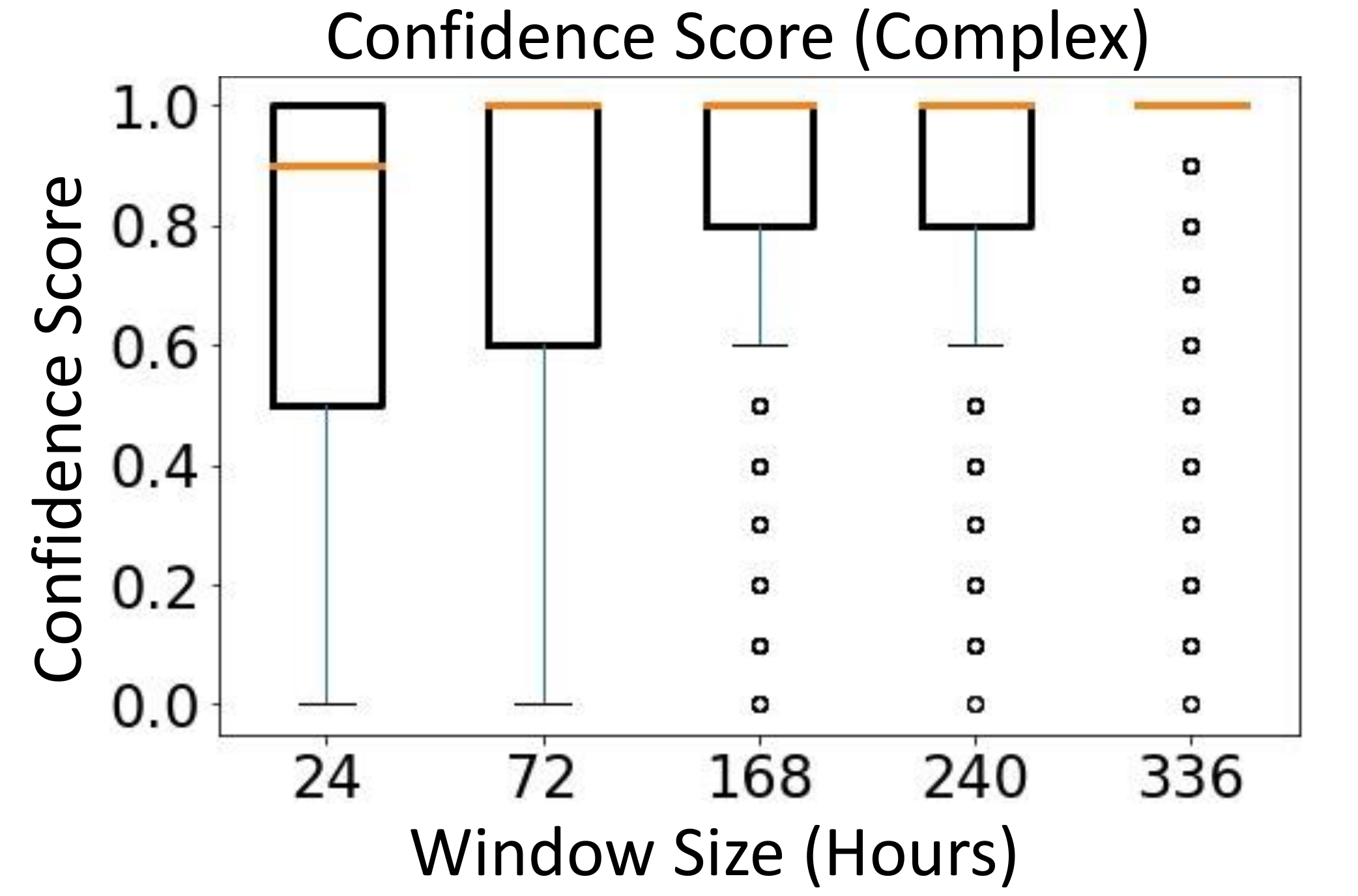}
% 		\caption{For the SKU recommended based on 30-day data.}
		% 	\end{subfigure}
% 	\vspace{-6mm}
	\caption{Confidence score distribution for the SKU recommended based on 30-day data.}\label{fig:AccuracyScoreDist}
	\vspace{-5mm}
\end{figure}

Since the SKU recommendation generated with \sysname is highly sensitive to the time period in which the performance counters are collected, a confidence score is surfaced along with the optimal SKU choice to provide the customer with some additional guarantees. We use the confidence score to encourage greater data collection, 
as Figure~\ref{fig:AccuracyScoreDist}
% as Figures~~\ref{fig:ConfidenceScoreDist} and \ref{fig:AccuracyScoreDist}
shows higher confidence associated with SKU recommendations that are made on performance input collected over a longer time span. %(which is more accurate). 
% Specifically, Figures~\ref{fig:AccuracyScoreDist} shows the distribution of the confidence score for the recommended SKU by assessing the performance footprint over a 30-days data and highlights that for customers with complex \pp curves, the confidence score is low when the window size is small (e.g., 1-week or less). On the other hand, for customers with simple (or flat) \pp curves, the recommendation SKUs from \sysname are typically associated with high confidence scores.
We examined confidence scores for successfully migrated customers in which we have at least 30 days' worth of performance data. As shown in Figure~\ref{fig:AccuracyScoreDist} by the various bootstrap window sizes tested, the confidence score values shift up as the time window of data collection increases past the 1-week interval. Preliminary results suggest that 1-week is the minimum duration needed to capture the variability in perf data required for a reasonable SKU recommendation. %for a representative workload. 
We thus encourage new migration customers to run the DMA tool for at least seven days.

\vspace{-1mm}

\subsubsection{Customers with Changing SKUs}\label{sec:changingSKU}
\begin{figure}[t]
	\centering
	\includegraphics[width=0.5\columnwidth]{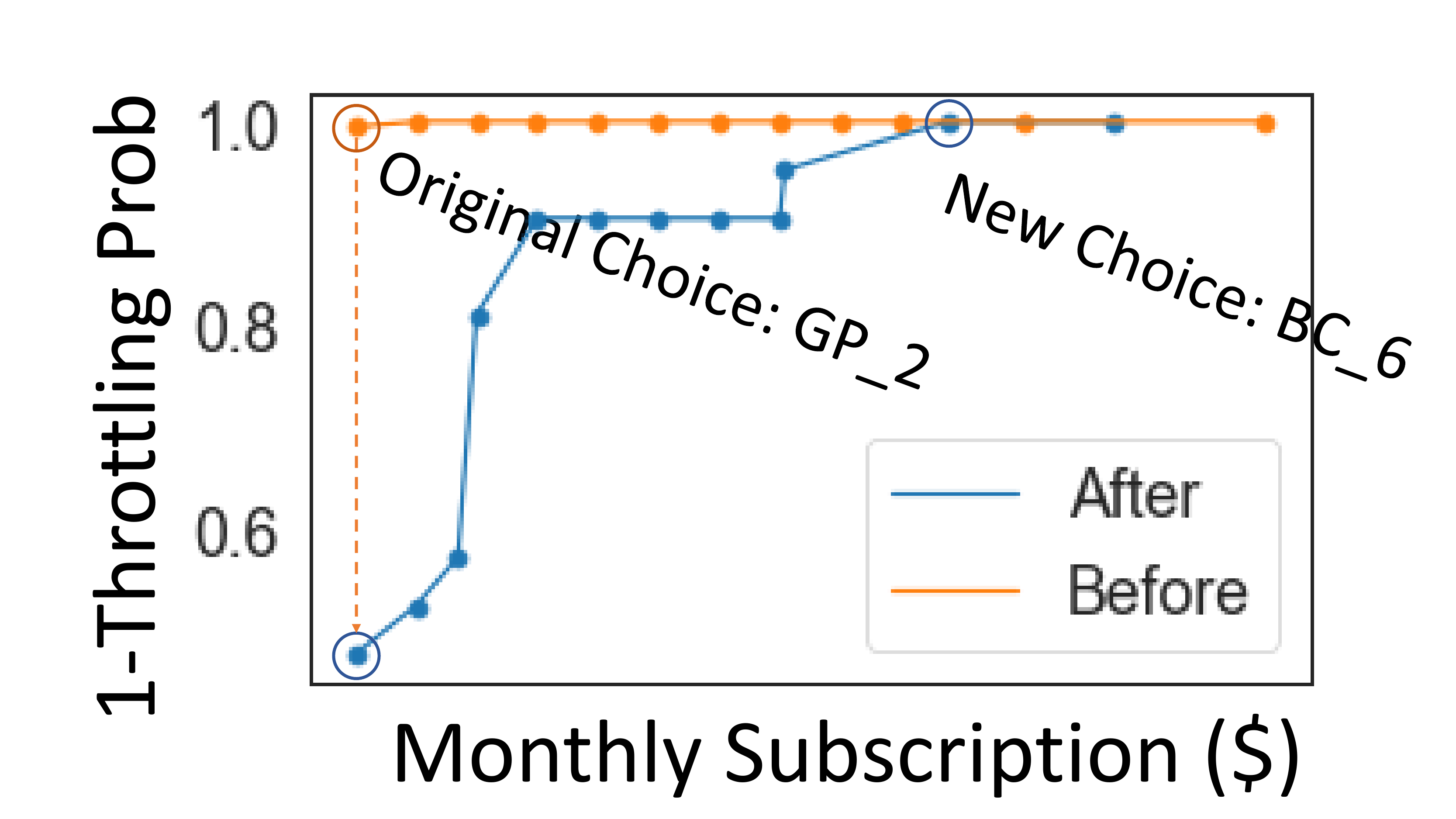}
 	\vspace{-3mm}
	\caption{Example of a set of \pp curves before (dotted line) and after (solid line) a SKU change.
		%		The customer originally fixed the SKU to Azure SQL DB GP vCore 2 on the dotted \pp curve; however retaining this SKU with the new workload demands as highlighted by the solid \pp curve is insufficient. To meet utilization needs at >90\%, the customer changed their SKU to Azure SQL BC vCore 6.
	}\label{fig:changing_SKU}
 	\vspace{-6mm}
\end{figure}

%Given the significant proportion of SQL DB and MI customers that do not have a definitive elbow point, 
Given the ease with which customers can change their allocated Azure SQL PaaS, we study successfully migrated SQL DB customers that made one SKU change between June 2020 and March 2021. This results in a subset of 77 customers that either upgrade or downgraded their initial SKU choice. For this set of performance histories, we study the \pp curves generated by \sysname \textit{before} the change and \textit{after} the change to understand whether the \pp curve is able to pick up on the changing customer resource utilization needs. 
%Given the flexibility that customers can change SKU of their Azure SQL resource, we observed about 10\% of our current customers constantly upgrade/downgrade their SKUs. We leveraged their resource usage data to better understand how they have changed their SKU decisions over time to validate our method.
%First, we examine a subset of customers that exhibited one change in their SKU selection over the past 40 days. We formed a set of ``before'' and ``after'' \pp curves for this subset of 77 Azure SQL DB customers; namely, a ``before'' \pp curve based on the resource profiles prior to the SKU change, and an ``after'' \pp curve based on the resource profiles after the SKU change.
As shown in Figure~\ref{fig:changing_SKU}, \pp curves adapt to changes in resource usage and can detect the need to change SKUs. For the particular customer workload highlighted here, the customer initially was using SQL DB GP 2 cores, but switched to SQL DB BC 6 cores. \sysname is able to pick up the need for this change as shown by the \pp curves generated before (dotted line) and after (solid line) the transition. If the customer had stuck to the original SKU choice of GP 2 cores, they would experience significant throttling (>40\%). The new SKU of BC 6 cores meets the customers' resource needs at 100\%. Since changes in resource utilization patterns trigger changes in the \pp curves, \sysname can automatically detect the need to change SKUs to accommodate changing workload requirements.

%and the previous selected SKU (GP with 2 vCores) was no longer adequate (it would have only met the customers' utilization less than 60\% of the time with the new \pp curve in blue).
%in fulfilling the customers' new workload demand. Had the customer in Figure~\ref{fig:changing_SKU} retained the old GP vCore 2 SKU, it would have only met the customers' utilization less than 60\% of time. 
%In this example, while there appear to be several SKUs by the blue dots on the solid \pp curve that meet the customers' needs at >90\% utilization, this particular customer chose the cheapest SKU that met 100\% utilization (BC with 6 vCores). Further study with domain experts of this subset of 77 customers that made a SKU change in the past 40 days revealed that customers who choose SKUs below 100\% utilization are primarily willing to negotiate on IOPS and Log Rate. And those SKU preference changes can be automatically detected by the \sysname customer profiler accurately.

\subsection{Data from On-Prem Workloads: Comparison with Baseline Strategy}\label{sec:baseline}
% \subsection{Comparison with baseline approach}
%\yz{questionable on this part, if we can say the one that is different and we can provide better recommendation? 13 small samples, no SKU recommendation is bad one major improvement.}

% \ww{Once we attain access the AzMigrate clusters, we would need to run our method against a number of other methods like the baseline, previous DS, hardware mapping / lift-and-shift approach, other methods in the literature, and demo how our SKU recommendation is the "optimal" in some way ... we can use metrics like annual cost savings, what the customer is actually using in the cloud now (whether they have changed the decision over time), and maybe results from Chengcheng's resource profiler? this section requires a lot more work, but i think it might be needed in solidifying this approach from a reviewers' standpoint}

% For on-prem customer data, we compared \sysname with the baseline approach (with the default aggregation of 95th percentile).
Since there is no objective ground truth label for the ``optimal'' SKU for new migration customers that have not yet migrated their workloads to the cloud, we compare the SKU recommendations generated from \sysname against the recommendations generated from the baseline strategy. As the baseline strategy simply reduces each perf dimension into one scalar (max) quantile value, we expect the SKU recommendations from this naive approach to be less optimal. For comparative purposes, the baseline strategy is set to reduce the time series vector to the value that corresponds to the 95\% percentile. Since the majority of performance histories were extracted from relatively idle workloads, we focus on three real customers whose perf history would allow for a robust SKU recommendation. For this limited set, we identified 10 instances in which \sysname is able to provide a more appropriate SKU recommendation. More specifically, 80\% of the time \sysname recommends a SKU that can actually meet customers' workload latency requirements, while the baseline incorrectly specifies a lower-end SKU. For the rest of the cases, the baseline strategy actually fails to provide any SKU recommendation as it assumes that no SKU can meet the requirement of all resource dimensions at 100\%. It is in these particular scenarios that \sysname is especially useful, as it allows customers to negotiate on various perf dimensions to select relevant SKUs. Further investigations are ongoing to compare how the baseline SKU recommendations against that of the \sysname strategy.
\subsection{Synthesized Workload}\label{sec:syn}

As workload replay is still considered the best practice when it comes to validating whether a new SKU can handle a specific workloads' resource needs, we verify \sysname with this strategy. Given that we want to operate within the confines of certain data privacy restrictions, in that we want to be able to make SKU recommendations without accessing customer data/queries, we leverage a tool that synthesizes new workloads solely based on the customers' performance history. While still a work in progress, this tool has demonstrated success at reconstructing new workloads that resemble real customer workloads simply by taking as input historical performance traces. The synthesized workload is generated by combining pieces of standardized benchmarks (e.g., TPC-C~\cite{tpcc}, TPC-DS~\cite{tpcds}, TPC-H~\cite{tpch}, and YCSB~\cite{ycsb}) with different database sizes (i.e., scaling factors), query frequency, and concurrency (number of concurrent clients). When executed on the same machine as that of the original workload, the performance traces of these synthesized workloads mimic that of the original. In this section, we validate our SKU recommendation approach by executing synthesized workloads on a selected number of relevant SKUs. 

%To better validate the SKU recommendation engine ongoing work is to prototype a workload synthesizer to demonstrate the utility of leveraging such performance counters as a proxy for customer workloads and reconstruct new workloads that resemble the customer workloads.  The workload can vary by time to capture the dynamic nature of customer workload. We derived the \pp curve for 4 SKUs that are tested (hosted on 4 different Azure Virtual Machines as in 

\begin{table}[!t]
	\caption{ SKUs used to execute synthetic workloads}
	\label{tab:hardware}
	\vspace{-0.1in}
	\small
	\begin{tabular}{lccccc}
		\toprule
		ID     & vCPU & Memory & Cache  & Throughput                        & Disk                      \\ \midrule
		SKU1  & 4 cores    & 16 GB  & 100 GB & 6000 IOPs                         &                           \\ 
		SKU2 & 8 cores   & 32 GB  & 200 GB & 12000 IOPs                    &                           \\ 
		SKU3& 16 cores    & 64 GB  & 400 GB & 154000 IOPs   &                           \\ 
		SKU4 & 32 cores   & 128 GB & 800 GB & 308000 IOPs                    & \multirow{-4}{*}{2TB SSD} \\ \bottomrule
	\end{tabular}
	\vspace{-0.05in}
\end{table}
% \begin{table}[!t]
% 	\small
% 	\begin{tabular}{lccccc}
% 		\toprule
% 		ID     & vCPU & Memory & Cache  & Throughput                        & Disk                      \\ \midrule
% 		 D4s v3  SKU1  & 4 cores    & 16 GB  & 100 GB & 6000 IOPs                         &                           \\ 
% 		 D8s\_v3 SKU2 & 8 cores   & 32 GB  & 200 GB & 12000 IOPs                    &                           \\ 
% 		D16as\_v4 SKU3& 16 cores    & 64 GB  & 400 GB & 154000 IOPs   &                           \\ 
% 		 D32s\_v4 SKU4 & 32 cores   & 128 GB & 800 GB & 308000 IOPs                    & \multirow{-4}{*}{2TB SSD} \\ \bottomrule
% 	\end{tabular}
% 	\caption{ Hardware platforms used to execute synthetic workloads \textmd{(ID refers to machine type provided by Azure Cloud)}~\cite{stitcher}}
% 	\label{tab:hardware}
% % 	\vspace{-0.45in}
% \end{table}

\begin{figure}[t]
	\centering
	\includegraphics[width=0.6\columnwidth]{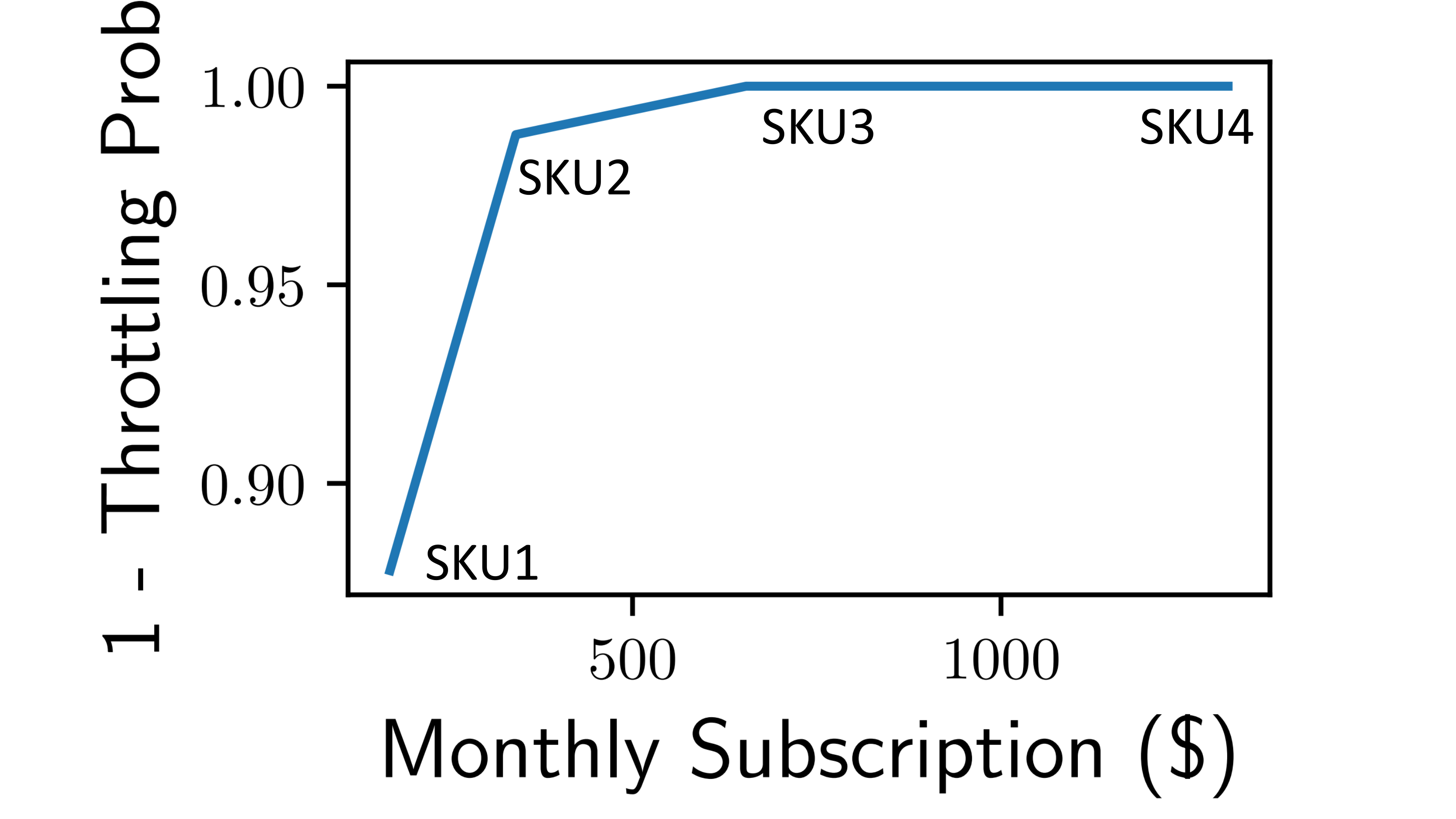}
	\vspace{-4mm}
	\caption{The \pp curve for the synthesized workload generated on relevant SKUs.
		%		The customer originally fixed the SKU to Azure SQL DB GP vCore 2 on the dotted \pp curve; however retaining this SKU with the new workload demands as highlighted by the solid \pp curve is insufficient. To meet utilization needs at >90\%, the customer changed their SKU to Azure SQL BC vCore 6.
	}\label{fig:stitcher_ppl}
	\vspace{-3mm}
\end{figure}

\begin{figure}
	\centering
	%	\begin{subfigure}[b]{\columnwidth}
	%		\centering
	%		\includegraphics[width=.5\columnwidth]{images/cpu_peaks.png}
	%		\caption{CPU utilization over an assessment period.}
	%	\end{subfigure}
	%
	%	\begin{subfigure}[b]{\columnwidth}
	%		\centering
	%		\includegraphics[width=.5\columnwidth]{images/cpu_peak_pp.png}
	%		\caption{Price-performance curve based on CPU utilization.}
	%	\end{subfigure}
	\centering
	\begin{minipage}[b]{0.5\columnwidth}
		\includegraphics[width=\textwidth]{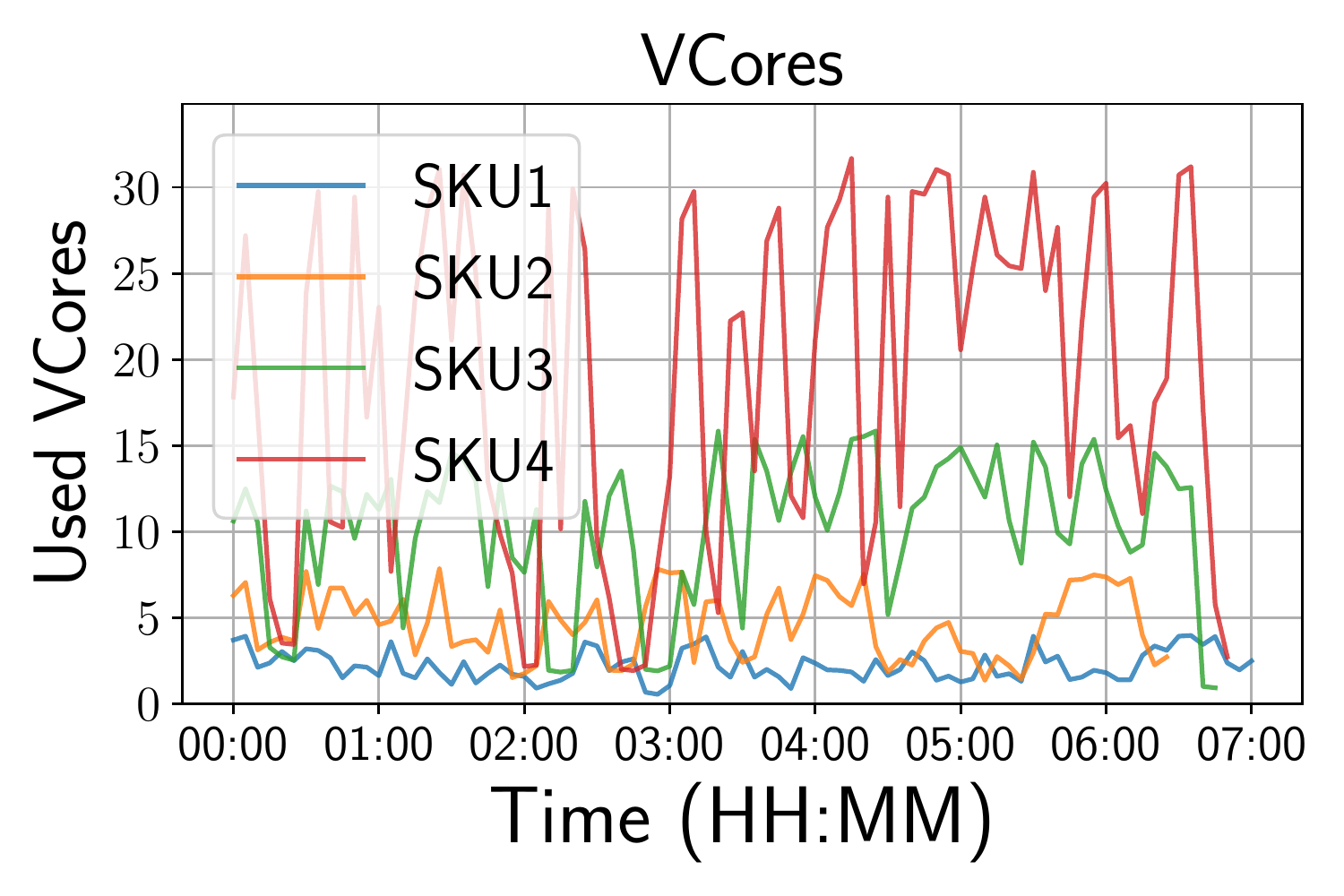}
	\end{minipage}%
	\begin{minipage}[b]{0.5\columnwidth}
		\includegraphics[width=\textwidth]{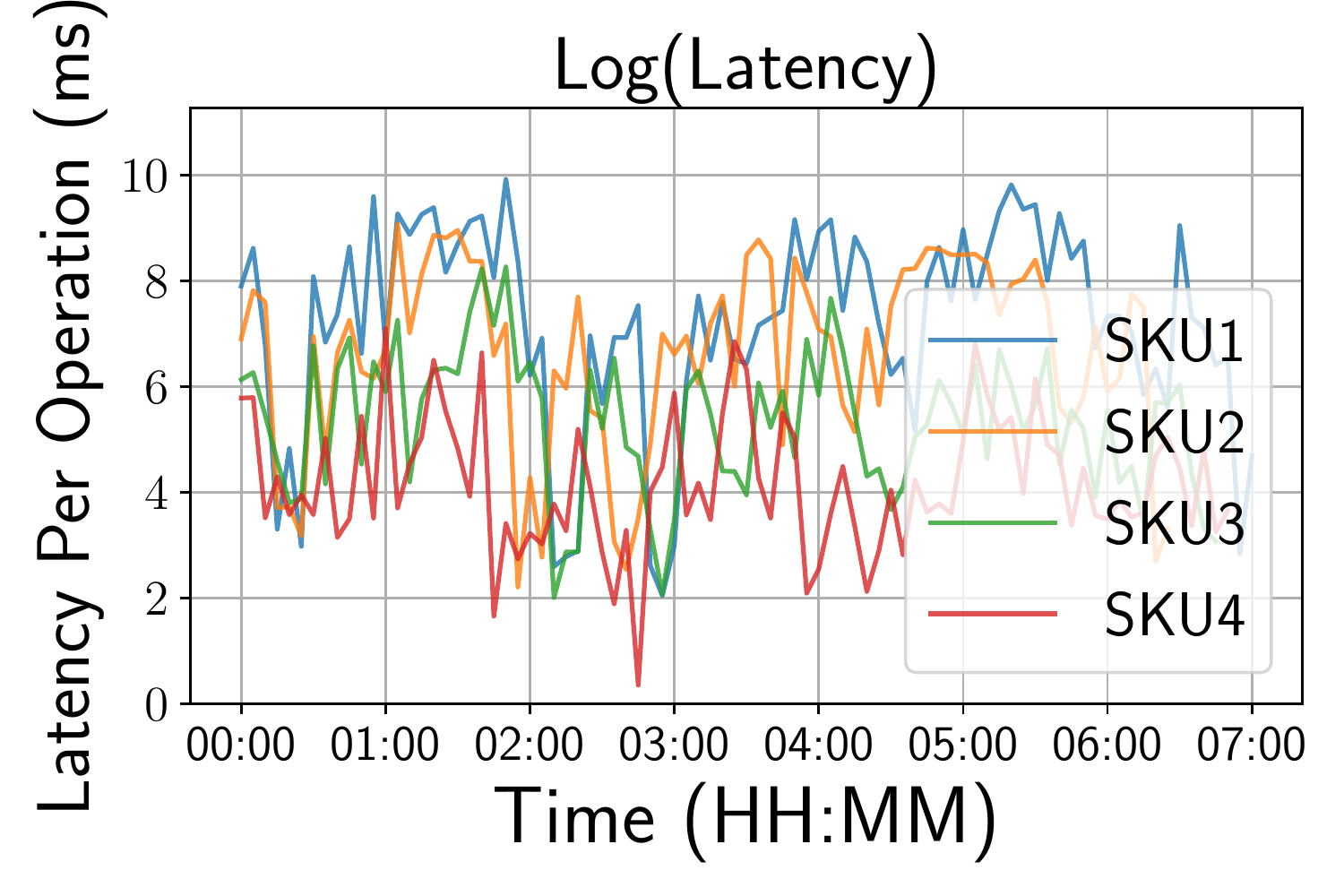}
	\end{minipage}%
	
	\begin{minipage}[b]{0.5\columnwidth}
		\includegraphics[width=\textwidth]{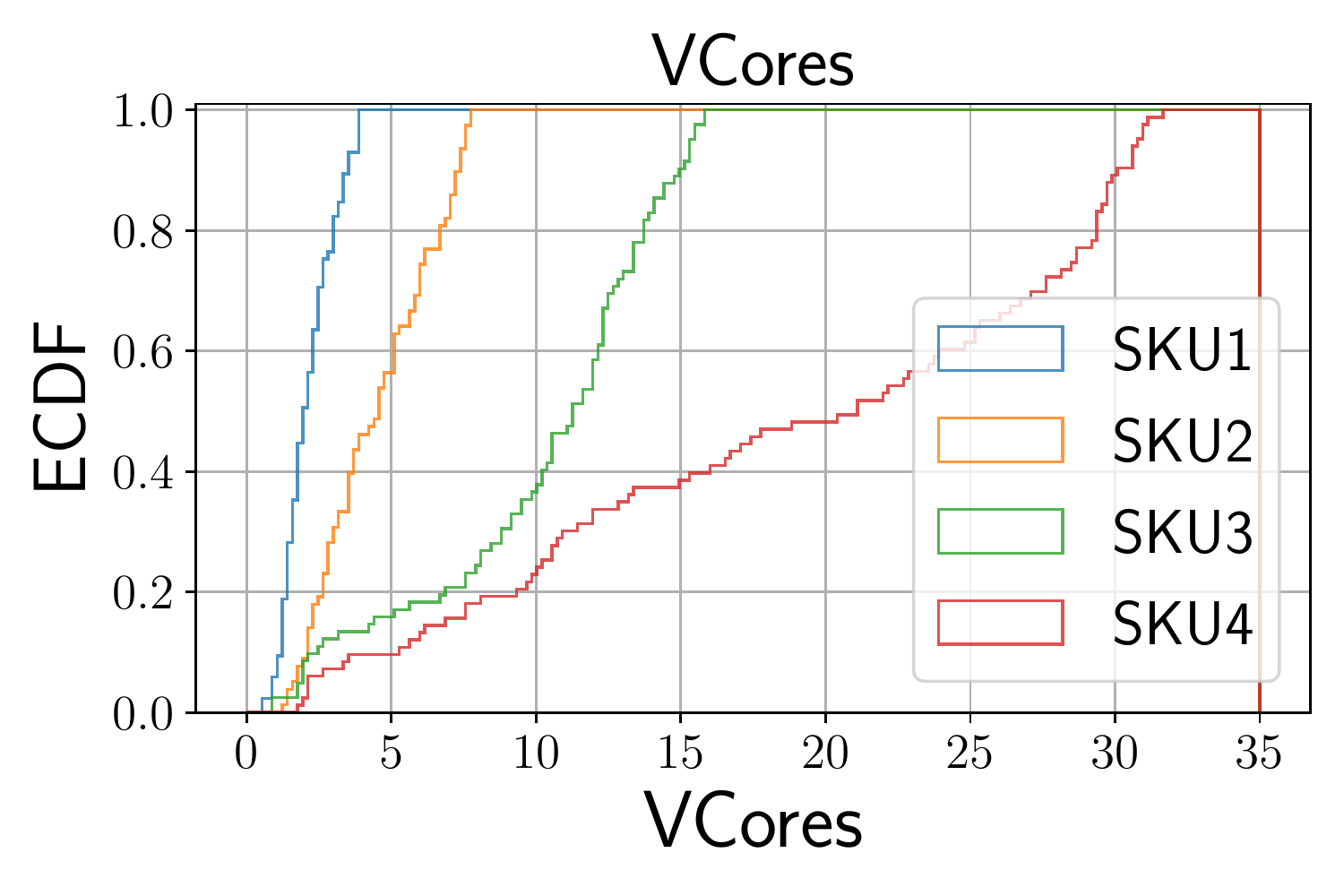}
	\end{minipage}%
	\begin{minipage}[b]{0.5\columnwidth}
		\includegraphics[width=\textwidth]{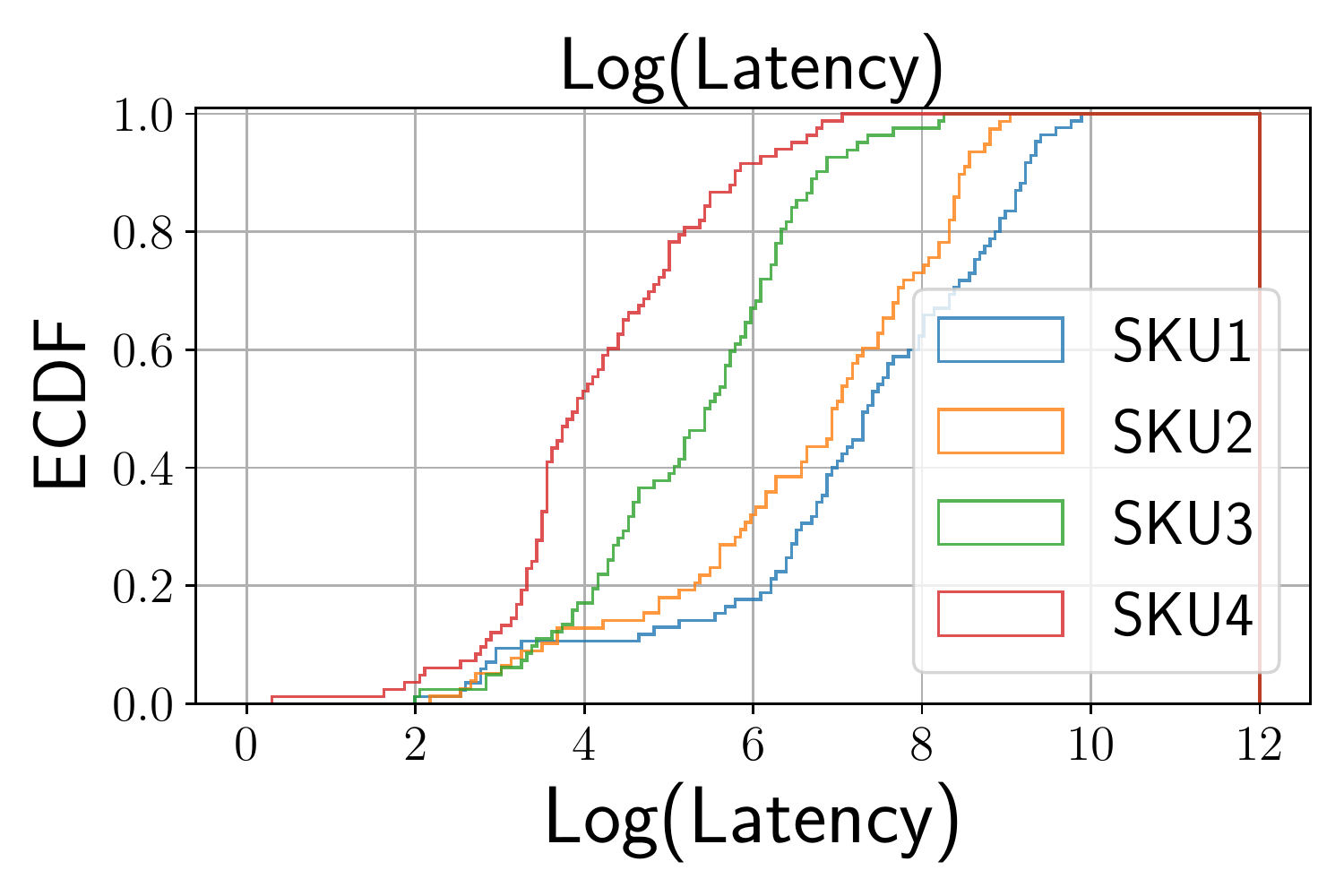}
	\end{minipage}%
	\vspace{-2mm}
	\caption{Performance counters when the synthesized workload is executed on four different Azure SQL SKUs.
		\vspace{-1mm}
		%	Their subsequent \pp curve (right) automatically adjusts for these anomalies as highlighted by the fact that Azure SQL SKUs that have fewer than 16 cores still satisfies this customers' needs with greater than 90\% frequency.
	}\label{fig:counter}
\end{figure}

Figure~\ref{fig:stitcher_ppl} shows one example of a \pp curve generated from the performance footprint from a synthesized workload. 
Using \sysname, the optimal SKU is identified as SKU2, details of which are outlined in Table~\ref{tab:hardware}. 
To validate the accuracy of this SKU recommendation, we replay the synthesized workload on the four relevant SKUs ranked in this \pp curve. Since there is no perf counter that directly measures ``throttling'', we examine the suitability of the four SKUs via the CPU and latency performance traces. As shown in Figure~\ref{fig:counter}, we can see that the lower-end SKUs results in an increase in IO latency. This is expected as this specific SKU has fewer cores, and with heavy workloads, latency should increase as the resource hits performance bottlenecks (e.g., the customer might experience this as throttling). SKU2, however, is appropriate as latency is within the range that the customer is comfortable with. Additional discussion with our solution engineers verifies that for this synthesized workload, SKU2 is the most cost-efficient option without sacrificing too much on performance. When executing this workload on the cheaper SKU option, SKU1, the workload is severely throttled, which further supports the SKU recommended by \sysname. While our work here is anecdotal, the execution of several synthesized workloads results in the same outcome as illustrated in Figures~\ref{fig:stitcher_ppl} and \ref{fig:counter}.

%and the optimal SKU can be identified. When replaying the same synthetic workload on 4 different SKUs, we collect the performance counters in the same dimensions (e.g., CPU, Memory, IO Latency, and LogRate). Figure~\ref{fig:counter} shows the CPU and IO Latency.

%We can observe that with a lower-end SKU, the IO Latency increased. \cc{Such differences appear on heavy workloads, where a higher-end SKU could afford more computing resources, e.g., vCores. Accordingly, a lower-end SKU inevitably reaches performance bottleneck and suffers performance degradations----its data processing capability cannot keep up with network bandwidth and user queries.}
%\yz{Per discussion with the domain experts, we determined that SKUX is the optimal one due to limited amount of time with resource throttling and relatively cheaper price. When executing the workload, by observing the other program's running on the same server, we also found that the machine with SKUX executing such workload is severely throttled, which is consistent with our expectation.}

%\begin{figure}[t]
%	\centering
%	\includegraphics[width=1\columnwidth]{images/db_resource_usage_corr}
%	\caption{The Correlation Heatmap of Resource Usage}
%	\label{fig:corr}
%\end{figure}

\subsection{Discussion}
%\change{Based on the experiences of field engineers utilizing .. } \cmt{customer as the direct user most of the time.}
Evaluation of our elastic strategy with data from existing cloud customers and from new migration customers, that have exclusively run their workloads in on-premise data platforms, demonstrates the ability of \sysname to provide quick, personalized SKU recommendations with high accuracy. More specifically, \sysname is able to generate the exact same SKU choice as successfully migrated customers have chosen (and fixed for at least 40 days) for 89.4\% and 96.7\% of SQL DB and SQL MI customers, respectively. Since this result excludes customers that are highly over-provisioned, we believe the accuracy we achieve with \sysname is a good proxy for the true accuracy of the best SKU. Future work includes adding a feedback loop in which we can re-train our clustering step based on customer satisfaction with their allocated SKU. From the perspective of our solution architects, few of our customers are under-provisioned. 

Our experimentation also revealed that approximately 10\% of existing SQL PaaS customers are over-provisioned, initiating a key program in which over-provisioned customers are right-sized to help them realize significant cost savings over time. \sysname has also been successfully deployed and used in the field as its recommendations can be clearly explained by the \pp curves. This framework can also be easily extended to include additional perf dimensions. 

\sysname is also designed to be plugged into any migration assessment infrastructure to enable other migration scenarios. Efforts are underway to integrate \sysname into a broader total cost of ownership (TCO) project, in which customers moving to Azure would be able to systematically compare the differences between keeping their workloads on-prem, moving to a hybrid cloud, or transferring workloads to Google Cloud Platform (GCP)~\cite{gcp}, Amazon Web Service (AWS)~\cite{aws}, and/or Azure. \sysname plays an important role in the TCO application in ensuring customers get an accurate picture of what the optimal SKUs and what their costs will be.

\section{Related work}\label{sec:related}
% Many tools exist to assist the process of selecting the right cloud target.To distinguish our automated SKU recommendation engine, \sysname, from existing tools, we describe a range of migration support systems that are also model-driven.
We exclude discussion around the efforts related to workload replay (e.g.,~\cite{galanis2008oracle, yagoub2008oracle,kopaneli2015model}) or simulation (e.g.,~\cite{saez2015performance}) due to the practical constraints imposed by lack of direct access to user data and cloud billing costs limits.

Several surveys \citep{gholami2016cloud, razavian2015systematic, jamshidi2013cloud, khadka2013legacy, lane2011process, wei2010service} highlight the complexity of this decision-making point and cite tools like the Cloud Target Selection (CTS)~\cite{kopaneli2015model} and Cloudle~\cite{kang2010cloudle} that support migrating legacy applications to the cloud. CTS is a large question catalogue that aids new migration customers in choosing the right provider and cloud SKU; Cloudle is similar, but also allows customers to specify additional functional, technical and cost requirements. The REMICS process \citep{sadovykh2011remics} introduces an array of model-driven solutions to support migration of legacy systems, which is more extensive than survey-like approaches (e.g., CTS) and search engine-type strategies (e.g., Cloudle). It starts with an analysis of the legacy system including its source code, binaries, configuration files, and execution traces, and extends to other knowledge discovery methods that can be translated to cover aspects of the business process, rules, components and implementation. Thus, the migration process is much more hands-on and customized in adapting existing cloud services to that of the customer. However, those tools still require significant user input therefore are difficult to scale and hard to use. 

\section{Conclusion}\label{sec:conclude}

%Consider the mutability of cloud environments..
% Considering these aspects, we propose a model-driven solution that automates the process of identifying features that are important to consider in allocating the right SKU to a customer. 

%Migrating customers to the cloud is a complicated decision process. 
%%We investigated these approaches to better understand how we can develop a more generic, scalable solution and how we can best help our customers migrate their on-prem SQL instances to the cloud. 
%%While our current process entails identifying a one-to-one mapping, or replicating an existing on-prem server in the cloud, this often results in over-allocated resources. 
%To avoid oversimplification, and to reduce the overhead associated with the decision-making involved in the migration process, we developed \sysname as a more holistic means for migration; thus, we not only automate the process of information extraction to make (SKU) recommendations, but also automate the decision itself, guiding the user towards one optimal cloud choice. To the best of our knowledge, \sysname is the first system in production to propose this end-to-end solution in the migration context. 
%

%Given the 8 billion TAM for migrating on-premise workloads to the cloud,
Given the TAM for migrating on-premise workloads to the cloud is almost \$50B and growing continuously, there is a need for tools that can make quick, accurate and personalized SKU recommendations. The current de facto standard of manual investigations in order to map workloads to appropriate cloud targets is too time-consuming and difficult to scale. We introduce \sysname as a means to ease one key bottleneck of migration, automating the SKU decision point. This automation is possible because of our novel end-to-end SKU recommendation framework, based on traditional \pp methodology that makes recommendations solely on input from customers' performance footprint. This strategy has been back-tested on customer data that have successfully migrated to Azure SQL PaaS, and has subsequently been integrated into production with the release of DMA v5.5 in October 2021.

%The need to develop a tool to ease the migration process and encourage more customers to migrate remains an open challenge.
%We focus on making accurate, personalized and interpretable Azure SQL DB and MI SKU recommendations to help increase customers propensity to migrate and to reduce post-migration friction.
%We developed \sysname as a more holistic means for migration; thus, we not only automate the process of information extraction to make (SKU) recommendations, but also automate the decision itself to guild the users towards one optimal cloud choice that is comparable or even better than human judgment, which usually involves time-consuming manual investigation.
%We introduce a novel end-to-end SKU recommendation framework based on traditional price-performance methodology to make recommendations for migrating on-prem customers to Azure SQL PaaS. This approach has been integrated into production, in the existing DMA tool, and has been released with version 5.5 on Oct $9^{th}$ 2021.

Since \sysname has demonstrated significant improvement over the baseline algorithm for making Azure SQL DB and MI recommendations, work is currently underway to extend this approach to assess other offerings like Azure SQL serverless \citep{serverless}, hyperscale \citep{hyperscale}, IaaS (e.g., Azure SQL VM \citep{sqlvm}) and other database systems (e.g., Oracle~\cite{oracle}). 
%Advantages of extending \sysname to these offerings include the fact that Azure SQL serverless can automatically scale up and down compute based on workload demands and Azure SQL hyperscale can further augment this flexibility by providing highly scalable storage.
%In addition to these SQL offerings, our framework can be generalized and applied to other products and services. 
One concrete example is our engagement with Azure Data Factory (ADF) \citep{adf}, in which \sysname has been adapted to recommend appropriate compute infrastructure optimized by cost and performance.
Given the simplicity of the \sysname framework, we believe it can be applied to address many product challenges in which understanding the trade-offs between cost and performance is needed. 
% Price performance curves can become a more versatile tool for recommendation engines in general, and in circumstances in which there are multiple objectives that customers seek to optimize.
%%
%Accurate and personalized SKU recommendation increases their propensity to migrate to the cloud and can reduce the post-migration friction due to wrong sizing.
%%

%team to igure out how to apply our approach to recommend the compute infrastructure that is cost and performance optimized for customers given operation type, data size etc.
%There is a wide collection of literature on ranking algorithms in academia, as well as a number of industry best practices on developing price-performance curves; however, we have not yet found one that uses our particular approach in both domains.

\begin{acks}
We thank Jack Li, Zeljko Arsic, Aleksandar Ivanovic, Kirby Repko, Venkata Raj Pochiraju, Ajay Jagannathan, Drazen Sumic, Debbi Lyons, Randy Thurman for their workload insights. We thank Yuanyuan Tian, Anja Gruenheid, Rathijit Sen and VLDB reviewers for their feedback.

%The authors thank Kirby Repko, Venkata Raj Pochiraju, Ajay Jagannathan, Drazen Sumic, Debbi Lyons, Randy Thurman for their customer insights. Their experiences motivated our approach towards automatic SKU recommendation. We also would like to thank Jack Li, Zeljko Arsic, Aleksandar Ivanovic from SQL DB and MI engineering on their insights on customer workload performance. We thank Yuanyuan Tian, Anja Gruenheid, Rathijit Sen and VLDB reviewers for their feedback.
% This work was supported by the [...] Research Fund of [...] (Number [...]). Additional funding was provided by [...] and [...]. We also thank [...] for contributing [...].
\end{acks}

% \section*{Acknowledgements}
% The authors thank Kirby Repko, Venkata Raj Pochiraju, Ajay Jagannathan, Drazen Sumic, Debbi Lyons, Randy Thurman for their insights working with migration customers. Their experiences motivated our approach towards automatic SKU recommendation.  We also would like to thank Jack Li, Zeljko Arsic, Aleksandar Ivanovic from SQL DB and MI engineering on their insights of customer workload performance. We also thank Yuanyuan Tian, Anja Gruenheid, Rathijit Sen and VLDB reviewers for their feedback.

%\section*{Acknowledgements}
%The authors thank Randy Thurman and Venkata Raj Pochiraju for their insights working with migration customers in the field. Their
%experiences motivated our approach towards automatic SKU recommendation. We also would like to thank Jack Li, Zeljko Arsic, Aleksandar Ivanovic from SQL DB and MI engineering on their insights of customer workload performance. We also thank Yuanyuan Tian, Anja Gruenheid, and VLDB reviewers for their feedback.

%\begin{acks}
% This work was supported by the [...] Research Fund of [...] (Number [...]). Additional funding was provided by [...] and [...]. We also thank [...] for contributing [...].
%\end{acks}

%\clearpage
\balance
\bibliographystyle{ACM-Reference-Format}
\bibliography{sample}

\end{document}